\documentclass[10pt]{article}    

\usepackage{amssymb,latexsym,amsmath} 
\usepackage[mathscr]{eucal}  
\newfont{\sffl}{msbm10 at 16pt} 
\newfont{\sff}{msbm10 at 10pt}

\begin{document}           
\title{\vskip -.36in Dirichlet integral dual-access collocation-kernel space\\
          analytic interpolation for unit disks: DIDACKS I\thanks{\small{Approved for public  release.}}
}

\author{Alan Rufty\\         
\\
P.O. Box 711\\
Dahlgren, VA. 22448}
\maketitle                 

\newcommand{\KD}{K_{\text{D}}}

\begin{abstract}
This article considers, along with its various direct extensions and ramifications, a technique for analytic interpolation over the exterior of a unit disk using complex poles in the interior---as well as corresponding techniques for the exterior of a real unit disk and for the interior of a real and complex unit disk.  More specifically, a simple closed-form technique is presented for approximating analytic functions in the exterior of a complex unit disk by linear combinations of simple poles located in the disk's interior at distinct points.  This is accomplished by developing special kernel spaces labeled dual-access collocation-kernel spaces.  In particular, it is shown that for two common norm settings a bounded point replicating kernel exists with the ubiquitous form $1/(z - z_k)$, where $z$ is in the exterior region of interest and where the fixed source point $z_k$ is inside the disk.  When the $z_k$'s are specified, both norm settings yield exact closed-form equation sets for the associated pole strengths.  The two norm settings are the standard inner product one for functions prescribed on the unit circle (where boundary value differences are minimized) and the Dirichlet integral one (where field energy differences are minimized over the whole exterior of the unit disk).  
Besides $1/(z - z_k)$, logarithmic point source kernels are also considered in both the real and complex planes.  Higher order poles are also considered.  Results of numerical tests are given.  Certain educational possibilities are addressed.  Relationships to the Cauchy integral formula, Bergman kernel theory and Szeg\H{o} kernel theory are also addressed.  For one of the norm settings---Dirichlet-integral dual-access collocation space (DIDACKS)---this is the first of several planed articles and these other articles will address various generalizations and applications of the basic technique presented here.
\end{abstract}


\newcommand{\SubSec}[1]{

\vskip .18in
\noindent
\underline{{#1}}
\vskip .08in

}
\newcommand{\ls}{\vphantom{\big)}} 
\newcommand{\lsm}{\!\vphantom{\big)}} 

\newcommand{\eq}{:=}

\newcommand{\smallindex}[1]{\text{\raisebox {1.5pt} {${}_{#1}$}}}

\newcommand{\mLarge}[1]{\text{\begin{Large} $#1$ \end{Large}}}
\newcommand{\mSmall}[1]{\text{\begin{footnotesize} $#1$ \end{footnotesize}}}

\newcommand{\Blbrac}{\rlap{\bigg{\lceil}}\bigg{\lfloor}} 
\newcommand{\Brbrac}{\bigg{\rbrack}} 

\newcommand{\Dr}{\mathscr{D}_r}
\newcommand{\R}[1]{${\mbox{\sff R}}^{#1}$} 
\newcommand{\mR}[1]{{\mbox{\sff R}}^{#1}}
\newcommand{\RR}{${\mbox{\sff R}}$}
\newcommand{\mRR}{{\mbox{\sff R}}}
\newcommand{\C}{${\mbox{\sff C}}$}
\newcommand{\mC}{{\mbox{\sff C}}}
\newcommand{\D}{{\text{D}}}
\newcommand{\sps}{0}

\vskip .05in
\noindent
\begin{itemize}
\item[\ \ ] \small{\textbf{Key words:} {unit disk, complex poles, potential theory, interpolation, Dirichlet form, inverse problem,\\ \phantom{Key words. L}  Laplace's equation, point collocation,} fundamental solutions, point sources, multipole}
\item[\ \ ] \small{\textbf{AMS subject classification (2000):} {Primary 30Exx. Secondary 31Axx, 65E05, 78A30, 9702}}
\end{itemize}

\renewcommand {\baselinestretch}{1.35} 
 
\section{Introduction}\label{S:intro}

  First consider the paper's content and structure.  This article develops a simple closed-form interpolation approach that can be used to approximate complex analytic functions in the exterior of a unit disk by a finite set of spatially distributed poles in the disk's interior.    This approach replicates values of the analytic function to be fit (and/or its derivatives) exactly at specified points and thus requires only limited information at a finite number of selected exterior points.  Furthermore, this closed-form point replication approach also minimizes the difference between the given function to be approximated and the approximating form (i.e., the modeling difference) for one of two common norms of interest and it produces a set of $N_k$ exact closed-form linear equations that can easily be solved for the corresponding $N_k$ pole or point source strengths.  The dual interpolation and approximation nature of the solutions obtained from these equation sets means that they are actually  interpolation fits.  Mathematically this is accomplished by developing special kernel spaces that, while closely related to reproducing kernel Hilbert spaces, are different from them.  (Notice that since it is assumed that noting is known about the function to be modeled in the interior of a unit disk, the strength of the individual poles cannot be found in the usual way by performing interior line integrals and then applying the theory of residues.)  The general kernel structures associated with these spaces are briefly outlined in Section~\ref{S:KernelSetting} and various associated notational and nomenclatural conventions are given in Section~\ref{S:notation}.  This is followed up with a discussion of the approximation space setting in Section~{\ref{S:PreDIDACKS}.  Although the corresponding theory for $\mathbb{R}^2$ functions is not explicitly considered until Section~\ref{S:DIDACKS}, in order to properly set the stage for their introduction the foundation for their use is also laid in Section~{\ref{S:PreDIDACKS}; hence, there are two primary classes of functions that must be dealt with here: (a) A restricted class of $\mathbb{C}$ functions that are analytic over the exterior of a complex unit disk. (b) A restricted class of $\mathbb{R}^2$ functions that are harmonic over the exterior of a real unit disk.  These two classes of functions are defined in such a way as to be coextensive (i.e., in direct correspondence).  In the end, this is done by requiring that the class of admissible analytic functions be restricted to those that have a power series representation of the form (\ref{E:series1}) and then by assuming that corresponding $\mathbb{R}^2$ functions can, in principle, be obtained through a process of harmonic completion.   While other ways of defining the relevant classes of admissible functions exist, one reason that this approach is used is that it reduces the level of the core material (basically the material from Section~{\ref{S:SIDACKS}} on) to a bare minimum.  The accessibility of this core material, in conjunction with certain other factors, opens up several unique educational possibilities that are partially outlined below and in Appendix~A.  Section~\ref{S:KernelSetting} starts out by considering the class (b) functions just mentioned, but defined over a more general domain, which corresponds to a real functional analysis setting with the functions defined over some sub-domain (and possibly an unbounded exterior region) of $\mathbb{R}^2$.  In Section~\ref{S:KernelSetting} it is shown that the direct analytic/real function correspondence mentioned above side-steps standard Hilbert space and Sobolev space theory in favor of a structured (i.e., augmented) pre-Hilbert space setting.  As partially discussed in Section~\ref{S:PreDIDACKS}, this has the effect of  circumventing certain issues related to the kernel forms used, as well as other issues tied to matters of physical interpretation.  Section~\ref{S:PreDIDACKS} also addresses various other foundational topics such as the uniqueness of simple pole expansions (and real dipole expansions) and the possibility of interpolation fit solutions that have good modeling properties while being the solution of an ill-conditioned equation set.  At the end of Section~{\ref{S:PreDIDACKS} a transition is made to  the more concrete mathematical matters that frame the core of the article.

 Section~{\ref{S:SIDACKS}} gives a straightforward and self-contained example of these kernel structures within the context of the first norm setting.  This first norm minimizes the modeling differences on the unit circle and is formally proportional to the standard Hilbert space $L^2$ norm of Lebesgue square integrable functions over this circle.   This norm is labeled the standard norm so as to distinguish it from this regular $L^2$ norm and its attending Hilbert space backdrop.  The space associated with this first norm is called the standard-integral dual-access collocation-kernel space (SIDACKS).  (Since acronyms are common throughout the paper a list of them is included in Section~\ref{S:notation} for the reader's convenience.)  Connections to the Cauchy integral formula, Bergman kernel theory and Szeg\H{o} kernel theory are also addressed in Section~{\ref{S:SIDACKS}}.  The second norm is introduced in Section~\ref{S:DIDACKS} and it is based on the well known Dirichlet integral, which means that it is an energy norm for the modeling difference.  The associated space for this inner product structure is called a Dirichlet integral dual-access collocation space (DIDACKS) and the primary mathematical relationships for it are also presented in Section~{\ref{S:DIDACKS}}.

   Up to Section~\ref{S:RealPlane} the article is primarily concerned with  analytic functions in and on the exterior of a complex unit disk (i.e., $f(z)$ for $z \in \mC$ and $|z| \geq 1$).  As noted above, the general perspective adopted here is that problems in \C\ and \R2 can largely be considered interchangeable since: (a) a function that is harmonic over some \R2 region of interest can be uniquely completed to form an analytic function $f$ over the same complex region (c.f., Definition \ref{S:RealPlane}.1} and surrounding discussion) and (b) both the real and complex components of an analytic function $f$ can separately be considered \R2 harmonic functions (where the complex component is usually simply ignored).  Thus the technique presented in the paper also forms a basis for framing associated harmonic fitting problems in \R2 where logarithmic sources commonly occur.  Section~{\ref{S:RealPlane}} first extends the technique to handle complex logarithmic point sources and then these complex plane results are carried over to the real plane, \R2; hence, closed-form equation sets are shown to exist for the strength of logarithmic (or dipole) point sources in both norm settings. (In Section~{\ref{S:RealPlane}} it is pointed out that it is actually more efficient to map the norm structures themselves from \C\ to \R2 rather than use the above harmonic to analytic mapping and then the return mapping.)  Next extensive, but not exhaustive, numerical testing is performed in Section~{\ref{S:examples}}.  Given the point replicating nature of the algorithms here, these tests serve not only to check the software implementation, but also, to a certain extent, the formalism itself.  As an aside it is worth noting that educators looking for interesting examples of ill-conditioned matrices should consider the results presented in Section~\ref{S:examples}, but by the same token those who are inexperienced should exercise a great deal of caution when dealing with systems that have condition numbers as large as those that are considered in Section~{\ref{S:examples}}---expecially if any computational or measurement errors are present.

  While the focus on the exterior of a unit disk is retained up through Section~\ref{S:examples}, in the next and final section (Section~\ref{S:IntDACKS}) it is shown that the basic approach can also be developed for interior approximation problems, where the region of interest is the interior of a unit disk and the approximating poles are located outside the unit disk.  This means that all of the various embellishments to the formalism that were developed in the prior sections can also be carried over to this interior setting.  Appendix~A contains additional discussions of certain side issues: In addition to the educational factors mentioned above, Appendix~A also considers certain pedagogical aspects of the material in the main body of the article.  Finally, Appendix~B briefly addresses the side issue in the complex setting of what specific expression to consider for the Dirichlet integral when the functions are either analytic or antianalytic.  Before proceeding readers may wish to note that kernels are generally placed on the left and that the complex conjugate of the first factor is taken in an inner product rather than the second---in Section~\ref{S:notation} the acronym left conjugate form (LCF) is introduced to define this convention.

   Next several issues pertaining to the paper's notation and style will be addressed, but  first, however, it is useful to briefly consider the paper's origins.  The method described herein had its genesis in a relatively recent detailed reexamination of gravity modeling linear least squares (LLSQ) and non-linear least squares (NLLSQ) techniques that the author first developed and successfully applied about a quarter of a century ago at the Naval Surface Weapons Center Dahlgren Division and, as such, the method should be considered technology spin-off.  As noted in the last line of the abstract, the associated general interpolation/modeling framework, as well as certain applications settings and theoretical implications, will be described in a sequence of articles, of which this is the first.  Most of the articles in this sequence are already largely complete.  Even though they are labeled as a sequence, most of these articles should be more-or-less independent of the others and so, except for a few exceptions, can be read out of sequence.  Later in this sequence a brief discussion of the general goals of this sequence is planned.  After the core articles in this cycle are complete an overall survey article will be written that may contain this discussion of goals, provided that time and circumstances permit.  Finally the author wishes to thank his group leader, Larry D. Wilkerson, and his branch head, Howell J. Boyles, for their patience and forbearance during the gestation of this sequence of articles and, in advance, for their review or perusal of each of the articles in the sequence.  As usual, any lapses of judgment with regards to content and style, as well as the occurrence of typos, are the sole responsibility of the author.  Due to the overall accessibility of these articles, since mathematical errors or typos can be easily detected from the surrounding context and since DIDACKS implemented algorithms display an underlying point collocation (or interpolation) property, which serves as a self-consistency check, typos are not considered to be a significant issue in themselves, but by the same token a reasonable effort will be undertaken to eliminate them.

  Having considered it structure, content and origins it is useful to examine the article's overall style.  The detailed discussions that ensue from this examination will occupy the rest of this section.  Due to educational and other factors, in part of the discussions that follow  certain topics that are not normally addressed in research articles are addressed and this results in some encroachments of the nominal research article style conventions.  To otherwise limit the extent of these stylistic encroachments, these discussions occur here primarily only in the rest of this section and in the appendices of this and future articles.

  Given these, as well as other style and notational issues, the reader may wonder why the paper was written as it is.   First, while it may not be clear to the reader from the content of this article alone, a somewhat irregularly shaped hole was perceived by the author to exist in the literature that entailed pure, applied and educational factors and, after attempting to balance all of these aspects, the approach initiated here seemed to be the most appropriate way to undertake an initial effort at partially filling this hole.  Second, either the material here is sufficient to stand on its own merits or it is not.  If it is, then it can endure these stylistic abuses.  If it cannot, it will be ignored in any case.  Aside from the overview article dealing with goals, the general stylistic tack taken in the other articles in the sequence will be that of a nominal research journal article where it is entirely up to each individual reader to discern whether the material in each article does or does not have intrinsic merit with regards to his or her own framework of research or educational interests.  While this nominal journal style could have greatly simplified matters here, it would have been necessary to abandon all educational factors (in the end, these factors often do not take care of themselves so they must be taken into account at the onset).  Those readers who are not otherwise interested in these larger issues---namely educational aspects, aspects dealing with the overall setting of the approach and an attempt at assessing the novelty of the technique---may simply skip the rest of this section (as well as Appendix~A) and proceed directly to Section~\ref{S:KernelSetting}.

  With regard to the issue of novelty or uniqueness of the approach, which is discussed at length at the end of this section, in the end it seems premature to make a definitive assessment here and it is the hope that the passage of time will allow for a better perspective.  At present what seems to be mainly lacking, aside from those connections pointed out in the text, are known primary references---if any such references exist at all.  Note that while there may be some truth in the unthinking reflex reaction of some that anything of interest in basic elementary complex variables that can be done has already been done, this applies only to the extent that this well known material is looked at from a more-or-less standard perspective; however, if this standard material is looked at from a sufficiently different point of view new results may indeed be possible.  What seems to be different here is to take the idea of a dual-access collocation kernel (DACK) and its ramifications seriously.  At the very least, this concept seems to shed some new light on otherwise well known results.  It also introduces a coherence and unity into otherwise disconnected results in complex variables and potential theory.  Finally, in some real sense it is this overall context that transforms a collection of interesting but isolated results into a potentially useful tool kit.  Setting forth this context is one of the overall goals of the sequence of articles.

  For better or worse this article can be viewed from either of four different perspectives:
\begin{enumerate}
\item[\{\ I\ \}]
As a research article. (Or more specifically one on analytic functions and the theory of poles and related topics in $\mathbb{R}^2$.)
\item[\{\,II\,\}]
As an article addressing educational aspects.
\item[\{III\}]
As a review article.
\item[\{IV\}]
As the first in a sequence of articles that introduce either an author's unique line of research or an author's unique take on existing material.
\end{enumerate}
It is not unusual for various articles to simultaneously display three of these attributes.  In fact, most research articles contain a brief introductory section that sets the new results in context by  introducing and briefly surveying the relevant research literature, thus encompassing attributes \{I\} and \{III\}.  (As noted below, the literature is not surveyed here since, aside from long established results that are discussed at length later, no closely related research results are at present known by the author to exist.)  By the same token, a research article that is the first in a sequence may quite often simultaneously encompass some mixture of attributes \{I\}, \{III\} and \{IV\}.  Likewise, the first of a sequence of educational articles might be expected to encompass aspects \{II\}, \{III\} and \{IV\}.  What is unusual here is that, to some degree, all four aspects are simultaneously displayed.  This is unusual since attributes  \{I\} and \{II\} are generally considered to be diametrically opposed.  Thus, the usual strategy that is employed by most researchers is to simply trump any significant educational (and review) possibilities and produce a pure research article.  As discussed below and in Appendix~A, this was not done here since these educational prospects are thought to be significant and unique here and the overall impact on the research aspects of the article that were incurred by taking these prospects into consideration was thought to be acceptable (but, of course, in the end this is a matter of personal taste and the same trade-offs will not necessarily be made even in the future articles of this sequence).  This does, however, lead to certain inevitable complications of style and notation and it is thus useful to consider these complications from each of the four perspectives just listed.

    \{I\} This article is primarily intended as a research article, but when viewed from the perspective of any one of these four article classes, there are several noteworthy complications here due to the concurrent educational and review aspects of the article.  First, the article is longer than it might otherwise have been, but in the main body of the article the extra steps that are included in some derivations, as well as the extra explanations, should lessen rather than increase the effort required by researchers to read and understand it. Otherwise, when the results are considered on their own terms, a fairly mathematically direct approach results.  Second, from the perspective of a research article on analytic functions and the theory of poles, there are many various ramifications and implications that are not brought up, so upon closer examination the article has an unsatisfactory open-ended nature to it.  (A few of these possibilities are addressed below, but most are simply by-passed.)  Third, also from the perspective of a research article on analytic function theory, some non-standard notation is used.  In particular, as previously noted one convention adopted here is that kernels are usually placed on the left and the complex conjugate of the first factor is taken in an inner product rather than the second factor.  The reasons for doing this are discussed at length in Section~\ref{S:notation}, where it is argued that here there are strong pedagogical advantages inherent in adopting this convention, which include directly leveraging analogies from linear algebra and operator theory, as is done, for example, in the theory of integral equations.
Introducing kernel based approaches quickly and painlessly is very important, not only for areas of pure mathematics, but for various diverse areas of applied science.  Aside from providing a minor nuisance to researchers, there is little real down-side here to adopting this inner-product conjugation convention, provided that the use of this convention is largely confined to an area that can be considered a new specialty or distinct sub-discipline, which is assumed to be the case here.  It is very important that conventions not be needlessly mixed in long established areas of mathematical research, since added complexity and confusion may otherwise result. 

  \{II\} The educational possibilities just alluded to are worth further elaboration, if even from a limited preliminary perspective since, as just noted, they have a bearing on the style and notation of the paper itself.  In fact, as discussed at the end of Appendix~A, the acronym DIDACK was chosen not only to imply direct connections to Dirichlet's integral, but to imply certain educational and self-educationally attributes of the basic approach---especially in the self-educational sense of an autodidact.  The other main educational possibilities are also briefly listed in Appendix~A, but any actual realizations in this area will have to depend upon the efforts of others since the author plans to make no systematic efforts in this direction.  With regard to educational factors then, the goal of this series of articles is merely to preserve the \underline{possibility} of educational aspects in case others wish to follow up on them.  Moreover, besides notation, the only systematic pedagogical concessions that will be made in future articles is that some effort will be made to keep them reasonably accessible. 

  Finally, there is a problem from an educational perspective with the article that is common to the three other perspectives also.  Specifically, in a graduate class room setting, it is incumbent on lecturers to point out what is standard known material and what is novel and unique---unique both with regards to what has been recently done by others and what the lecturer has added to it.  This is necessary since students are often at a loss to ascertain the boundary separating these different types of materials for themselves and one expectation of a core course in complex variables is to provide practitioners not only with an understanding of standard core concepts, but with a common shared language so that they can communicate with others in a professional capacity and understanding what others know or do not know is a key part of this communication process.  The problem here is that even researchers who are very experienced with the standard core material of complex variables may well vary on their assessment of what is actually new and unique here.  It is for this and other reasons that a fairly long discussion of this topic is included at the end of this section.

 \{III\} Next, consider problems with the article when it is viewed from the perspective of a review article.  First, and primarily, a review article of a mathematical field must be about some established area of research or sub-discipline and it generally should contain extensive references to the known work.  DIDACKS theory is not an established research area yet so there are no appropriate references; moreover, at present there is very much an open-ended quality to the research presented here.  Nevertheless, the material presented here forms a novel perspective and summary of much of the standard fare of complex variables, but if the reader's intent is simply to review complex variables, then there are many fine existing texts.

 \{IV\} Since this article was expressly written as the first in a sequence of research articles, it perhaps works best in this role, but there are still issues.  First, the overall goals are not addressed here.  Second, no attempt will be made here to indicate applications, but applications related issues will be addressed in future articles of this sequence.  Third, with regards to pure mathematics research, although absolutely no attempt will be made here to convince the reader that other actual additional research possibilities exist, there are three general classes of research areas where DIDACKS theory may, in some sense or other, be suggestive:

\begin{center}
\ 
\underline{Three General Types of Pure Mathematics Research Areas Worth Considering} 
\end{center}
\begin{itemize}
\item[(A)]
Well established and otherwise ``mined-out'' areas, which are worth re-examining from the perspective of the approach presented herein.
\item[(B)]
Currently active areas of research that may yield otherwise unexpected results when examined from  the perspective of the approach presented here, in the sense that DIDACKS theory may prove to be suggestive to an otherwise receptive researcher.
\item[(C)]
DIDACKS theory may, in some way or other, suggest entirely new lines of research.
\end{itemize}
To make all of this slightly less abstract, an example from each of these research areas that might be worth considering from a DIDACKS point of view is given next:
\begin{center}
\ 
\underline{An Example of Each Type of Research Area and DIDACKS}
\end{center}
\begin{itemize}
\item[(a)]
There are classic theorems in the theory of analytic functions and poles that link the value of an analytic function to its value at a sequence of points.  There are also theorems in approximation theory that give error bounds and that link the value of a function to a series of  partial sums of approximating functions.  DIDACKS theory deals with both of these classes of results and thus may imply that there are some unrecognized links between them.
\item[(b)]
When considered from the DIDACKS perspective, there may be interesting insights that emerge from analytic matrix polynomial theory.  (For classic references to analytic matrix polynomial theory see those in \cite{MatrixPoly1} and  \cite{MatrixPoly2}. \cite{MatrixPoly2} also gives a self-contained overview.)
\item[(c)]
 The concept of dual-access kernels and dual-access collocation kernels are introduced here and they may be worth exploring in their own right.
\end{itemize}

These particular examples are given primarily in the hopes that others will, at least, consider them in passing.  First, present plans are to not explicitly address either (a) or (b) further here or in any other articles in the sequel.  Also with regards to type (B) theories, there are many other approaches in analytic function theory that apparently have little in common with the approach taken here, such as Carleson interpolation and Pick interpolation (or Nevanlinna-Pick interpolation), but that are always worth examining from a different perspective due to their inherent richness and one way to do this would be to compare and contrast them with the DIDACKS approach outlined here.  (For example, with regards to Nevanlinna-Pick theory, in spite of overall differences, there are certain obvious research questions that stick out here and are worth exploring.)  Second, item (c) frames part of the discussion of kernel structures given in Section~\ref{S:KernelSetting} and may be partially addressed in future articles in the sequence; nevertheless, even item~(c) will not be, in any sense, be completely settled in this sequence of articles.

  Finally, the motivation behind item (c) is worth briefly considering.  To date, only symmetric reproducing kernels have received the full attention of the mathematical community.   The conventional wisdom seems to be that since an asymmetric kernel can generally be reformulated as a symmetric one and symmetric kernels have very nice properties only symmetric kernels are worth considering; however, specific counterexamples contained in this article indicate that this may not be a universally valid assessment.  The hope is that the discussion in Section~\ref{S:KernelSetting}, when coupled with other preliminary analysis and examples given in later articles in this sequence, will convince some researcher(s) that dual-access kernels and dual-access collocation kernels are  worth exploring in their own right.  The approach to this issue taken here is to try to place SIDACKS and DIDACKS theory within a larger nomenclatural context, which requires drawing certain distinctions that are not usually drawn---these distinctions are taken up in Section~\ref{S:KernelSetting}.  The nomenclatural complexity that ensues partially results from the fact that the perspective taken here is somewhat unique and partially from unavoidable conflicts with standard kernel naming conventions; however, as indicated above, these linguistic complications are offset by an easy overall accessibility of the article itself so that relevant implications can be readily drawn from the surrounding context.

\begin{center}
\ \\
\underline{\textbf{What is new here?}} 
\end{center}
Generally this is a question asked by readers who view mathematical progress as an exercise in taxidermy rather than as a growing dynamic enterprise and such readers are not usually interested in the possibility of a new synthesis or framework; nevertheless, this question is worth exploring in some detail at some point in this sequence of articles so this analysis is undertaken here.  Although this same question can be entertained again at various points in the future and for other DIDACKS contexts, the general points raised here should also be relevant in other settings as well so this type of analysis will not be repeated.  [Likewise, while the emphasis in this analysis is on the overall kernel setting, one can raise the same question with regards to various other specific (and obvious) concepts (such as the paired point formalism), which in some sense or another all appear to be new, but such items will not be discussed separately.]  The convention of research articles is followed here and in other articles in this sequence that assumes all results, when not otherwise noted, are new unless they are obviously known results.  In this connection, a concerted effort was made to search the relevant literature to ascertain significant precedents, but aside from several obvious connections to Szeg\H{o} and Bergman kernel theory that are discussed later in the text, there appeared to be none.  As discussed below, this is very surprising in itself, but it would be even more perplexing if the general approach developed in the sequel had occured somewhere or other in the literature and had been generally ignored, since this would otherwise seem to entail a significant oversight on the part of the broader mathematical community itself.  Although suitable references are not known, it does seem very likely that in the (mainstream?) literature the fact that Szeg\H{o} and Bergman kernels can be recast in the form $1/(z - z_k)$ has long been recognized; however, without the concept of a dual-access collocation kernel and the recognition of its general significance this fact was dismissed as a mere curiosity.  It would seem, then, that what is obviously new here is the overall perspective and with regards to particular results there is some question as to where the actual boundary lies between what is known and what is new.  However, even this assessment is not entirely clear, since there are at least six inherent difficulties in actually performing a complete and exhaustive search of all the relevant current and historical literature to ascertain precisely what is actually known and what is not; moreover, it is not even clear whether this assessment is too conservative or not conservative enough.

  First, there is the sheer magnitude of the current mathematical literature that would have to be searched.  For example, it might initially seem that only the complex setting (including analytic function theory and its various offshoots) would have to be searched since results that are known for real disk geometries are invariably carried over to the complex setting, but even this assumption need not be true since the work here is also tied to other broad areas of mathematics, including approximation theory, potential theory, integral equation theory and functional analysis (specifically reproducing kernel Hilbert space theory), which each have their own independent literature.  Second, a broad search of the current literature on analytic function theory and complex variables is more than sufficiently daunting, but matters are somewhat worse than usual here since most of the topics discussed realistically could, and perhaps should (and thus may), have been developed 50 years ago or much more.  This means that the literature to be searched must potentially span the entire historical record of a better part of a century.   Third, this search is compounded by the fact that various countries (including all of the major historically industrialized countries) have their own rich and varied traditions in complex variables and that the literature of each of these countries has their own associated technical and national languages.  Fourth, there are many separate disciplines within each of these countries that practice complex variables (i.e., applied mathematics, physics and various engineering subspecialties).  Fifth, the difficulty in assessing what is new here is also greatly compounded by the fact that many of the fundamental relationships obtained in the sequel are derived by well known elementary techniques.  [For example, the derivation of (\ref{E:sigg3}) uses several ubiquitous series representations along with the correspondence $z \rightarrow 1/z$; moreover, (\ref{E:sigg3}) itself is an alternative form of the Cauchy integral formula.] Clearly, when a result is obtained by elementary means it is tempting to assume that it is well known when, in fact, it may not be, but at any rate it clearly has the unfavorable by-product of greatly extending the range of literature that must be searched to include literature with a more elementary scope.  Sixth, on closer inspection the material presented here is somewhat broader than one might expect and this induces distinct literature assessments issues of its own.  For example, even for a well trained historian of mathematics there is usually a subjective element involved in determining all relevant precursors and their impacts on some broad topic in mathematics; moreover, it is generally recognized that the task of assessing origins is especially difficult and often fraught with potential controversy.  In summary, all that can be asserted in the end is that, besides Szeg\H{o} and Bergman kernel theory, no direct precedents were uncovered by the author in his literature searches. 

 Temporarily setting aside connections to Szeg\H{o} and Bergman kernel theory, the lack of obvious direct precedents in itself leads to a quandary.  First, given that the results presented here seem to have a fundamental bearing on the theory of poles, if something along very similar lines had been obtained sometime in the past and its significance had been overlooked by the broader pure and applied community, this in itself would have been very surprising---in fact, much more surprising than the fact that it had not been discovered at all.  A matter of judgment, however, does enter here though: if the reader perceives the content of the present article as being very specialized and of limited interest then since many disconnected specialized results exist in the literature it might be reasonable to assume that some significant precursors exist in the literature that are unknown to the author.  (In this regard, it is not unusual for specialized results to be perceived by a researcher as new when they were, in fact, originally done 50 or more years ago.)  But even here there is a problem since the judgment that some results are of limited interest must be almost universally held or else those who have a different opinion will integrate them into their own research efforts or expository writings. 

 In summary, it seems likely that there is a real quandary here since otherwise how could something so simple have been overlooked, but there is a simple resolution of this paradox.  To the extent that lessons from the history of science also apply to pure and applied mathematics---and there are a sufficient number of concrete examples from the history of mathematics to argue that this is indeed the case---when a new point-of-view emerges 
different questions are asked and entirely new results emerge.  It is the author's contention that the DIDACKS programme can be considered a fresh point-of-view, but the breadth and full extent of this point-of-view may not in any sense be apparent until this series of articles is complete. 

  Finally, it is also useful to look at the content of the present article with the express goal of placing some overall bounds on what the original content actually is here.  Thus to set a crude upper bound on the originality of the work here first consider the concrete links between the formalism presented here and Szeg\H{o} and Bergman kernel theory, which are discussed at length in later sections.  These links lead one to think that specific replication properties [such as (\ref{E:sigg3}) and (\ref{E:CollPropD})] should probably not be considered new, which immediately gives an upper bound on the possible originality of much of the work presented here.  Alternatively, observe that for \C\ and \R2 disk geometries Szeg\H{o}'s and Bergman's kernel are, in some sense, equivalent to Cauchy's integral formula [as are all similar integral results such as Poisson's integral or Green's theorem] and they are thus themselves derivative results when considered at least from a sufficiently basic elementary level; nevertheless, each of these kernels is of great separate interest in itself, not only for its basic properties, but for the copious related research that it has spawned.  In this sense, one  might argue that the true measure of originality and significance can only be made in historical hindsight.  [One  might argue that the recast form of these kernels given by (\ref{E:sigg3}) and (\ref{E:CollPropD}) have some small claim to distinctness thus calling this bound into question.]  For the lower bound consider the following.  In the $\mathbb{R}^2$ potential theory literature there seems to be no natural way of performing simple logarithmic or dipole fits and from this void alone one can argue that the approach initiated here helps to fill a theoretical hole.  Moreover, there seems to be no good elementary formalism for performing simple pole fits---which can be considered a natural extension of elementary complex variable theory.  From these deficiencies one can (tentatively) conclude that at least the overall approach presented here is in some sense new, or that at least within the context of these applications it is new, thus giving the desired bound in the other direction.

\section{Basic Kernel Setting}\label{S:KernelSetting}
  
  In order to frame the basic general kernel setting of the DIDACKS approach it is useful to compare and contrast it with standard reproducing kernel Hilbert space (RKHS) theory.  RKHS theory has numerous well known merits  \cite{Aronszajn,Mate,RKHSbook}.  It is useful to introduce RKHS conventions by way of a conceptualized example.   For real valued $L^2$ functions defined over the real interval $\,\mathcal{I} \eq \{ x \in [a,\,b]\,\mid\,a,\,b\ {\text{and}}\ x \in \mRR\ {\text{with}}\ a > b\,\,\}$ and with the inner product $(f,\,g) \eq \int_{\mathcal{I}} f(x)\,g(x)\ d\,x$, a reproducing kernel $K_R$ has the property that $(K_R(x,\,y),\,f(x)) = f(y)$, where $x,\,y \in \mathcal{I}$ and $K_R$ is assumed to be bounded and it is symmetric: $K_R(x,\,y) = K_R(y,\,x)$.   Reproducing kernels for other inner products and other subregions $\Omega$ of \C\ and \R{n} are defined analogously for $n \geq 2$.  These subregions $\Omega$ are simply the allowed domains for the functions of interest.  For a fixed subregion $\Omega$ and associated inner product it is easy to show that when a reproducing kernel exists it is unique. This uniqueness underlies the usual naming convention for RKHSs, which generally attach a specific mathematician's name to an associated reproducing kernel.  A compact and precise alternative description of reproducing kernels can be based on the Riesz representation theorem (\cite{BergmanSpaces2}, pp. 242-244 of Ch. 9).

  Two RKHS examples of particular import follow immediately as generalizations of the above example.  First, the class of reproducing kernels where the norm is defined as a  straightforward integral of $|f|^2$ over $\Omega$ are known as Bergman kernels.  Secondly, when the unit disk interior in \C\ is the region of interest and the norm is the integral over the unit circle of the square of the magnitude of some $L^2$ complex valued function, then the associated kernel is called the Szeg\H{o} kernel.  As noted elsewhere the Bergman kernels for the interior and exterior of a unit disk and the Szeg\H{o} kernel will prove to be of special interest since they are closely related to the kernels studied here. 

  When Bergman and Szeg\H{o} kernels are put in a general context one arrives at the left hand side of the following nomenclatural hierarchy structure:

\vspace{.11in plus 1in minus .04in}

\hspace{.3in} 
\framebox[5.2in][l]{\hspace{-.35in} 
\parbox{6in}
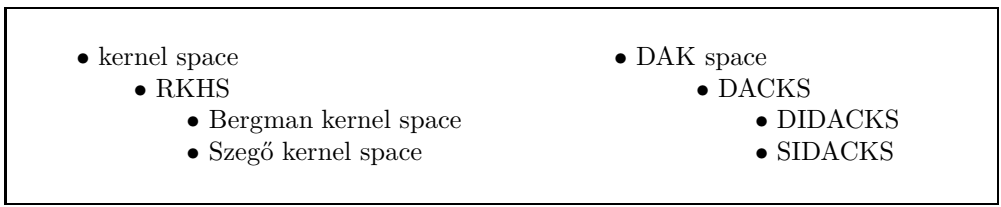
{
\begin{tabbing}
   \hspace{1.5cm} \textbullet\ ker\=nel s\=pace\hspace{5.0cm}\textbullet\ DAK\=\ spac\=e\\
           \>\textbullet\ RKHS                         \> \>\textbullet\ DACKS \\ 
                  \>\>\textbullet\ Bergman kernel space       \> \> \textbullet\ DIDACKS\\ 
                  \>\>\textbullet\ Szeg\H{o} kernel space         \> \> \textbullet\ SIDACKS     
\end{tabbing}
}
}

\vspace{.11in plus 1in minus .04in}


\noindent
which proceeds from general to particular on both the left and right hand sides.  Spaces of roughly the same generality are displayed on either side of each line.  The immediate goal of the remainder of this section is thus to explain the right hand side of this nomenclatural structure. 

  Before introducing the concept of a dual-access kernel (DAK) it is expedient to clarify the nature of allowable subregions.   While the applied parts of the paper from Section~{\ref{S:SIDACKS}} on center on disk geometries for $\Omega$ where $\Omega \subset \mC$ or $\Omega \subset \mR2$, in this section not only are more general geometries in \C\ and \R2 considered, but also often the cases $\Omega \subset \mR{n}$.  (Here and in the sequel \R{n} for finite $n \geq 2$ will be denoted simply by \R{n}.)  For subsequent developments it is expedient to impose additional structure on $\Omega$.  In particular, it is assumed that \C\ or \R{n} is partitioned into two regions: $\Omega$ and its compliment ${\Omega}'$, where at least one of these two regions is simply connected.  Further it is assumed that $\Omega$ contains the common boundary  separating these two regions (denoted $\partial \Omega$ here and assumed to exclude boundaries at infinity), so that ${\Omega}'$ is an open set.  The concept of a dual access kernel $K_D$ is then easy to define: it is a kernel with two arguments where one argument is in $\Omega$ and the other is in ${\Omega}'$: $K_D = K_D(x,\,q)$ where $x \in \Omega$ and $q \in {\Omega}'$.   A common example of a DAK arises in conjunction with the integral form of Poisson's equation for \R{n}, where the volume integral (or area integral in \R{2}) of a density function times a DAK integrated over a source region produces the potential at some point outside the source region.  When a DAK satisfies a replication property it is called dual-access collocation kernel (DACK) and the associated kernel space and inner product structure are known as a DACKS.  Clearly a standard reproducing kernel, such as a Bergman kernel, is not an example of a DACK, since both the kernel arguments of a DACK are not even in the same subregion, but in general the occurrence of a DACK is tied to that of some reproducing kernel.  A DACKS generalization of the reproducing property  exists and is known as a replication or generalized least squares collocation property.  For all currently known DACKs this replication condition can be expressed through the action of an associated one-to-one ``involution mapping'', $p(q)$,  that maps the region  ${\Omega}'$ into the interior of $\Omega$: $p(q) \in \Omega\setminus\partial \Omega$ for $q \in {\Omega}'$.  It also satisfies the property $p(p(q)) = q$.   The point $p$ is called the involution point.  For $q \in \mC$ this involution mapping becomes a ``conjugate involution mapping'' denoted $p^{*}(q)$ where $p^{*}$ is called the conjugate involution point.  It may seem circular to specify $K_D(x,\,q)$ with  $q \in {\Omega}'$ and then apply a mapping of the form $p(q) \in \Omega$ so that functions are finally evaluated in $\Omega$; however, this approach has certain innate advantages since it allows, among other things, for the natural consideration of inverse source problems.  As an example, consider the first DACK studied in the sequel; $1/(z - z_k)$, where $z_k$ is inside the complex unit disk and $z$ in outside this disk and the associated conjugate involution is point is $1/z^{*}_k$.  Obviously this kernel form links a given (simple) pole region or source region with an exterior region of interest---i.e., a field region.  In general, it is useful to limit the source region (i.e., the range space of $z_k$ here) to be a fixed proper closed subset of ${\Omega}'$ so that all the kernel forms studied are bounded. 

 There are some minor complications here compared to RKHS theory.  DACKS norms and kernels are no longer in general one-to-one correspondence with each other as they are in RKHS theory and a particular DACKS norm may have many different point replication kernels associated with it.  Likewise for a single fixed kernel, DACK replication properties may hold for several different norms that are all specified over the same given (field) region of space.  For example, as noted earlier, it will be shown that $1/(z - z_k)$ has two separate spaces or norms associated with it and the field region for both of these norms is the same.  This non-uniqueness causes some nomenclatural problems since it means that the standard RKHS naming convention mentioned previously cannot be followed.  This issue is circumvented here by simply specifying a norm structure along with the particular DACK of interest (which is generally understood from the context).  As one might gather from the discussion given in Section~\ref{S:intro}, many questions associated with overall DAK and DACK structures are largely open and one goal of this series of articles is to attempt to motivate others to explore these structures.  Much of the nomenclature just outlined is aimed at incrementally advancing this implicit goal under the assumption that labeling is a useful first step.  

   Given the potential connections between RKHSs and DACKSs outlined above, it is natural to wonder about the specific connections between well known reproducing kernels and the particular DACKs examples studied here.  At the end of Section~\ref{S:SIDACKS} a general connection will be shown to exist between the primary replication property for the first norm studied and Cauchy's integral formula.  It is well known that the reproducing property of Bergman and Szeg\H{o} kernels for the interior of a disk can be also be tied to Cauchy's integral formula.  Along similar lines it is fairly easy to show that for the exterior of a unit disk the primary DIDACKS replication property is linked to Cauchy's integral formula, as is the Bergman kernel reproducing property for the for the same region.  Obviously an indirect link exists between all these kernels.  Other direct connections are established in the sequel between DACKS kernels and the reproducing kernels just mentioned.  For the exterior problem at the end of  Section~\ref{S:SIDACKS} a direct link is shown to the interior Szeg\H{o} kernel. At the end of  Section~\ref{S:DIDACKS} it is shown that for the exterior of a unit disk the DIDACKS formalism can be linked to exterior Bergman space structures.  Finally, from the material in Section~\ref{S:IntDACKS} a direct connection between the interior form of the first norm and Szeg\H{o} kernel theory can be shown to exist.

\section{Notational and Allied Issues}\label{S:notation}

  This section addresses various notational issues and summarizes various acronyms and terms used elsewhere in the paper.  It is broken into various natural groupings where each item of interest is broken out separately.  Often the motivation behind a particular notation choice is given since choice of notation is important, but rarely discussed.

  The two conventions discussed first are not standard, but are adopted here since they have strong educational overtones.  These educational overtones that are briefly considered in context:
\begin{itemize}
\item
  For notational and conceptual clarity it is useful to be able to distinguish between the two normal uses of complex conjugation: (1) As an operator.  (2) To denote independent complex conjugate variables.  This distinction can be easily made since there are also two symbols in common usage: the superscript $*$ and overbar.  Because the matrix transpose operator is usually denoted by a superscript $T$, which is the convention adopted here, by analogy it is natural to use a superscript $*$ to denote the complex conjugate operator (i.e., $f^{*}$).   From a matrix perspective this usage is consistent and allows a dagger superscript to denote the Hermitian conjugate of a matrix; i.e., for Hermitian matrix $H$ it follows that $H^{\dag} \eq (H^{T})^{*} = (H^{*})^T = H$. This superscript $*$ is the default usage here, while an overbar is reserved for those rare occasions where there is a real need to introduce independent conjugate variables (i.e, $z$ and $\bar{z}$).   Most remaining ambiguities can be removed by the judicious insertion of parentheses [for example, see the right hand side of (\ref{E:sigg3})] where the issue of the independence of $z$ and $z^{*}$ does not arise (nor even need it be considered).  Further, any residual confusion that remains in specific functional forms can be eliminated simply by referring to the series form that is used to define the admissible class of functions (\ref{E:series1}). 

   Conjugate variables are useful and can occasionally greatly simplify an argument.  They are also widely used in the analytic function literature (such as in conjunction with Hardy and Bergman spaces).  The issue of the legitimacy of conjugate variables was brought up and settled long ago \cite{Nehari}, but the concept of conjugate variables is not generally covered in introductory books on complex variables.   By using both symbols in the way each is most naturally used, the full advantages of both notations can be accrued.

   Another approach often taken in the literature is to introduce the Wirtinger differential operators \cite{Wirtinger} commonly written as
\begin{equation}\label{Wirtinger}
 \frac{\partial\,}{\partial\,z} = \frac12\left(\frac{\partial\,}{\partial\,x} - i\, \frac{\partial\,}{\partial\,y}\right)\ \ {\text{and}}\ \ \frac{\partial\,}{\partial\,\bar{z}} = \frac12\left(\frac{\partial\,}{\partial\,x} + i\, \frac{\partial\,}{\partial\,y}\right)\,.
\end{equation}
Other labels associated with these operators are Wirtinger derivatives and the Wirtinger calculus.  Here the concept of an antianalytic function can be introduced to restore the symmetry between complex and complex conjugate variables and thus to fully leverage the innate advantage of these Wirtinger operators \cite{BergmanSpaces2}, but this approach is not taken here.  (The ``mirror symmetry'' of analytic and antianalytic function theory has a bearing on the next item under discussion, but this possibility is also not followed up in the sequel.)
 
 The reader who is an expert in analytic function theory may not see the utility of two separate conjugation symbols since the underpinning conceptual issues and tools are so obvious to him or her, but such a reader might want to consider the following hypothetical examples from matrix theory that suggests the sort of issues that a neophyte may encounter upon first exposure to analytic function theory when an overbar is used as an operator.  Specifically, suppose that an overbar is used in matrix theory to indicate transposition and that matrices that are functions of other matrices are considered:  $\mathbf{M}(\mathbf{N})$, for matrices $\mathbf{M}$ and $\mathbf{N}$.  Then what is the exact meaning, if any, of ${\mathbf{M}}(\overline{\mathbf{N}})$, of ${\mathbf{M}}\overline{(\mathbf{N})}$, of $\overline{\overline{\mathbf{M}}(\overline{\mathbf{N}})}$, of $\overline{\overline{\mathbf{M}}\overline{(\mathbf{N})}}$, of $\overline{\mathbf{M}(\overline{\mathbf{N}})}$, of $\overline{\mathbf{M}\overline{(\mathbf{N})}}$, of $\overline{\mathbf{M}}(\overline{\mathbf{N}})^{-1}$, of $\overline{\mathbf{M}}^{-1}(\overline{\mathbf{N}})$ or finally of $\overline{\mathbf{M}^{-1}}(\overline{\mathbf{N}})$.  Moreover, what is the exact difference, if any, of each of these expressions and can the reader easily interpret a handwritten version of these expressions?
That any effort whatsoever is required to answer these sort of questions indicates a possible inherent ambiguity in the notation itself.

Finally, the use of two separate conjugation symbols means that functions or expressions that are not analytic can be manipulated and handled just as naturally as ones that are.  Specifically, the resulting expressions for inner product evaluations need not necessarily be analytic.
\item
For pedagogical reasons, it is useful to distinguish between the two conventions that can be used to apply complex conjugation to the arguments of a complex valued inner product.  An inner product (or norm) where conjugation is applied to the first argument will said to be in left-conjugate form (LCF), while if conjugation is applied to the second factor it will said to be in right-conjugate form (RCF).  As an example, consider the complex form of the real inner product that was introduced in Section~\ref{S:KernelSetting} to explain reproducing kernels, where $\mathcal{I} \eq [a,\,b]$ is now taken to denote a line segment along the real axis of the complex plane.  The LCF of this inner product is $(f,\,g)_{LCF} \eq \int_{\mathcal{I}} (f(z))^{*}\,g(z)\ d\,x$, while the RCF is  $(f,\,g)_{RCF} \eq \int_{\mathcal{I}} f(z)\,(g(z))^{*}\ d\,x$.   Generally, when dealing with ``analytic functional analysis'' (i.e., complex RKHS theory, Bergman kernel theory, Szeg\H{o} kernel theory, $H^p$ theory etc.) it is the custom to use RCF inner product structures, which is the convention originally adopted by Bergman.  The LCF is adopted here for several specific reasons:
\begin{itemize}
   \item
   Here the resulting expressions from an LCF inner product application may not be analytic, but this does not matter in an applied setting. 
 \item
The linear equation sets that result from an LCF treatment are much more intuitive and easily derived [c.f., (\ref{E:TmuA})]. 
 \item
 LCF is the standard convention in quantum mechanics and elementary linear algebra.
\item
 LCF has also occasionally been adopted by mathematicians \cite{Korns} or used in concert with RCF by mathematicians \cite{MatrixPoly1}.
   \item
   From an educational perspective, the LCF convention allows much of the intuition gained in matrix theory to be directly leveraged.  Later, after the student is more comfortable with the concepts involved, the RCF can be introduced.  (It is also worth noting that in the past, compositions in the reverse order have been periodically tried in introductory group theory courses and other areas and have been generally found to be more confusing to students than the standard operator ordering, which they encounter during a first exposure to calculus and matrix theory.)  When a reproducing kernel, $K_R$, is considered as an operator it is natural to place it on the left in analogy with matrix theory; moreover, in the complex setting, at a certain level, it is largely a matter of choice as to whether the end result of $(K_R,\,f)$ or $(f,\,K_R)$ is required to be analytic (c.f, the concept of antianalytic functions mentioned above).
 \item
  From an applied perspective the choice of LCF or RCF is strictly a matter of convenience since either choice can be justified. Since matters are treated somewhat explicitly here, no inconvenience or confusion results by using LCF. 
 In either case some care is always required in using the resulting expressions (for example, see the above discussion of independent conjugate variables).
\item
 Obviously, it is easy to convert from LCF to RCF since $(g,\,f)_{RCF} \eq (g,\,f)^{*}_{LCF}$.
Further, while the RCF may be a real conceptual imposition to the neophyte or to the practitioner from another field, the conversion from one form to the other should pose no real problem for professional mathematicians---especially when one form or the other is maintained throughout a particular sub-discipline (as mentioned above, it is a working assumption here that DIDACKS theory can be treated as an independent sub-discipline).
\end{itemize}
With this notation, the real form of an inner product is given by $\frac12(g,\,f)_{RCF} + \frac12(g,\,f)_{LCF}$.  The paper's formalism was tested and checked in LCF.  In particular,  looking ahead notice that retracing the derivation leading up to (\ref{E:TmuA}) employing RCF yields ${\mathbf{\mu}}^T{\mathbf{T}}^T_{RCF} = {\mathbf{A}}^{\dag}_{RCF}$, where $\mathbf{T}_{RCF}$ denotes the matrix whose elements are $T_{{RCF}_{k',\,k}} \eq ({\mathcal{B}}_{k'},\,{\mathcal{B}}_k)_{RCF}$ and $\mathbf{A}_{RCF}$ the vector whose element are  $A_{{RCF}_{k}} \eq ({\mathcal{B}}_{k},\,f)_{RCF}\,$---which is equivalent to, but slightly more cumbersome than, (\ref{E:TmuA}) [and thus possibly confusing to a novice].
\end{itemize}
\underline{\ \ \ \ \ \ \ \ \ \ \ \ \ \ }\\
\\
In the sequel the labels real and complex setting are used in the following context:
\begin{itemize}
\item
The real setting $\implies$  the setting where the function to be approximated and the approximating function, as well as the basis functions that compose it, are real valued and have arguments in \R{n}.  The inner product is also real valued so that any scalar quantities involved are also real valued. 
\item
The complex setting $\implies$ the setting where the function to be approximated and the approximating function, as well as the basis functions that compose it, are complex valued and have arguments in \C.  The inner product is also complex valued and unless otherwise stated it is assumed to be in LCF.  It is also assumed that scalar quantities are also generally complex valued. 
\end{itemize}
\underline{\ \ \ \ \ \ \ \ \ \ \ \ \ \ }\\
\\
\noindent
  Before summarizing various acronyms, it is useful to discuss some of the index and subscript conventions:
\begin{itemize}
\item 
 The subscripts $k$ and $k'$ range over the values $1,\,2,\,3,\,\ldots,\,N_k\,$. 
\end{itemize}
 The subscript conventions associated with the major norms studied are:
\begin{itemize}
\item
A subscript $\sigma$ is used to denote the standard norm where $\sigma \implies$ the unit circle $\equiv$ the boundary of unit the disk ($\partial\Omega$) $\equiv$ the inner boundary of the exterior of the unit disk. (While the outward pointing normal for the interior and the exterior of the unit disk differ in sign, this normal is not explicitly introduced here for any of the standard-norm usages so this equivalence strictly holds.)
\item
Since the inner products associated with the Dirichlet integral assume different forms in the complex and real setting two different subscripts are used to distinguish these settings:
  \begin{itemize}
    \item 
     $D \implies$ complex setting
    \item
     $E \implies$ real \R2 setting.
  \end{itemize}
\end{itemize}
\underline{\ \ \ \ \ \ \ \ \ \ \ \ \ \ }\\
\\
There are also a number of standard acronyms worth summarizing here:
 \begin{itemize}
 \item
 LLSQ = linear least squares (when no confusion can arise this acronym is also used to denote a generalized linear least squares setting)
 \item 
 NLLSQ = non-linear least squares
 \item
 RKHS = reproducing kernel Hilbert space
 \item
 SVD = singular value decomposition.
 \end{itemize}
 Finally it is useful to summarize the main acronyms that are introduced in the paper:
 \begin{itemize}
 \item
 DACK = dual-access collocation kernel
 \item
 DACKS = dual-access collocation-kernel space
 \item
 DAK = dual-access kernel
 \item
 DIDACKS = Dirichlet integral DACKS (in either the real or complex setting)
 \item
 LCF = left-conjugate form
 \item
 RCF = right-conjugate form
 \item
 SIDACKS = standard integral DACKS (in either the real or complex setting).
\end{itemize}
Finally, as a linguistic aside, it is assumed herein that an acronym denotes an exact word sequence and thus, for example, that where it is appropriate the plural of an acronym can be formed by simply affixing a small letter s to the end of the acronym.  Also, although there is no established uniform convention in the literature with regards to forming the plural mathematical symbols, here and in future articles in the sequence an apostrophe s is invariably used in order to provide a useful visual clue as to where the math symbol itself ends (i.e, $a_k$'s is prefered over $a_k$s).

\section{Foundations of DIDACKS Theory}\label{S:PreDIDACKS}

 In what follows the basic strategy is the usual one of approximation theory, which involves introducing a norm and then minimizing the cost function
\begin{equation}\label{E:PHI}
 \Phi \eq \|f - \varphi\|^2 = \|f\|^2 - 2(f,\,\varphi) + \|\varphi\|^2\ \!,
\end{equation}
where $\varphi$ is the approximating function and $f$ is the function to be approximated.
When $\varphi$ is assumed to be a linear combination of fixed basis functions of the form
\begin{equation}\label{E:LLSQbasis}
\varphi(z) = \sum\limits_{k=1}^{N_k} {\mu}_k{\mathcal{B}}_k(z)
\end{equation}
a (generalized) LLSQ problem results, which is the main focus of this article.

  For the complex setting $z$, ${\mu}_k$ and ${\mathcal{B}}_k \in \mC$ and the inner product is assumed to be in LCF.  For the real setting $z \in \mR{n}$ and $f(z)$, ${\mathcal{B}}_k(z)$ and ${\mu}_k$ real valued. 

 For the real setting, the values of ${\mu}_k$ that minimize $\Phi = \Phi({\mu}_k)$ can be found simply by solving the linear equation set that results from setting the partials of $\Phi$ with respect to ${\mu}_{k'}$ to zero (and diving by two): 
\begin{equation}\label{E:LLSQ4}
\sum\limits_{k=1}^{N_k} ({\mathcal{B}}_{k'},\,{\mathcal{B}}_k){\mu}_k = ({\mathcal{B}}_{k'},\,f) \ ,
\end{equation}
which can be written more compactly as
\begin{equation}\label{E:TmuA}   
\boxed{\,\mathbf{T}\,\mathbf{\mu} = \mathbf{A}\,} 
\end{equation}
 where $\mathbf{T}$ denotes the matrix whose elements are $T_{k',\,k} \eq ({\mathcal{B}}_{k'},\,{\mathcal{B}}_k)$, $\mathbf{A}$ the vector whose element are  $A_{k} \eq ({\mathcal{B}}_{k},\,f)$ and $\mathbf{\mu}$ the vector whose elements are ${\mu}_k$.  Equation~(\ref{E:TmuA}) will prove to be pivotal in what follows and it occurs in different contexts at several places in the sequel.  [Here for the complex setting it is a simple matter to show that minimization of $\Phi = \Phi({\mu}_k)$ again leads to the same expression---see the steps leading up to (\ref{E:sigLLSQ}) or (\ref{E:DLLSQ}).]  

  The choice of inner product structure and of basis functions are of fundamental importance.
   As mentioned in Section~\ref{S:intro} the first class of basis function studied here will be ${\mathcal{B}}_k(z) = 1/(z - z_k)$, which is a DACKS kernel in the complex setting.  In the real \R2 setting this basis function corresponds to a dipole term.  In Section~\ref{S:intro} it was also as noted that logarithmic basis functions are DACKS kernels in both the complex and real \R2 setting.  

 Next, consider the Dirichlet integral and its implications here.  Although the approach in the core part of the article is to consider the complex setting first and then transition to the real setting, in order to gain a complete perspective the reverse course is taken in this section, hence the \R2 setting will be considered exclusively before transitioning to the complex setting.  Here a distinction is made between the Dirichlet integral and Dirichlet form---although it is not the usual custom in the literature.  For the real setting, the term ``Dirichlet integral'', denoted $\D[f,\,g]$, is taken to be an integral over a class of admissible functions that has a mathematical form proportional to the field energy (i.e., energy density) over the region of interest; i.e., (\ref{E:ReDirInt}) for the exterior of a unit disk in \R2.  For exterior disk regions the class of admissible functions here are essentially assumed to be bounded harmonic functions (with bounded partials of all orders $\eq C_B^{\infty}$) that tail off sufficiently fast at infinity. 
 When a function $F$, which is not assumed to be harmonic, satisfies much less restrictive conditions, the well known Dirichlet principle can be considered and the term ``Dirichlet form'' will be reserved for the associated form $\D[F,\,F]$ in this context.  Looking ahead to the complex setting, it is important to keep this distinction in mind since the complex form of Dirichlet's integral adopted here is derived under the explicit assumption that all admissible functions are analytic and, in this sense, it is not a Dirichlet form, nor is the question of the proper definition of a Dirichlet form in the complex setting even raised in the main body of the text.  (Although this is an intrinsically interesting question and it is worth observing that at the genesis of complex variable theory properties of analytic functions and other results were obtained using Dirichlet forms through the action of Dirichlet's principle.)

  Next, recall that the Dirichlet principle basically asserts that the minimization of the form $\D[F,\,F]$ over some specified interior region, subject to the constraint that $F$ match given Dirichlet boundary conditions, implies $F$ is harmonic.  Here for Dirichlet's principle to hold $F$ must satisfy certain minimal conditions (i.e., least restrictive conditions), as must the specified boundary conditions and the shape of the boundary itself.  Dirichlet's principle is also assumed to hold for exterior regions, but as discussed below there are a few more complications so this case is usually not addressed in the broader literature.  The historical counterexamples \cite{Courant} for interior regions, their associated history \cite{Mona}, the general nature of the requirements \cite{JostPDE,Rauch} and typical recent associated research efforts \cite{Jost} are of great interest in their own right, but are they are not especially relevant here since Dirichlet's principle is a tautology in DIDACKS theory---harmonicity is already assumed in the class of admissible functions itself.  Clearly there is a middle ground between these two stances where harmonicity is not assumed for the class of admissible functions, but the assumptions are strong enough on the class of admissible functions and the regions of interest so that Dirichlet's principle can be derived simply without untoward complications.  Historically assuming a certain class of explicitly defined admissible functions has been considered the acceptable way to attack Dirichlet's principle, with the first use in this context going back to Dirichlet himself \cite [p. 40]{Mona}.   Since the focus has generally been on proving Dirichlet's principle under the most general assumptions, these original attempts were flawed and counterexamples played a major role in uncovering the conceptual errors that were made at various historical junctures.  If the boundary of the domain under consideration is not pathological, many of the innate difficulties in a proof of Dirichlet's principle arise from the broad class of allowed functions that are inherent in the Hilbert space concept---especially completeness with regards to infinite sequences of functions.   Assuming a well formulated structured pre-Hilbert space setting by-passes these problems as well as problems in transforming Dirichlet's integral into an acceptable inner product structure.  For a structured pre-Hilbert space setting, it is first of all assumed that a pre-Hilbert space setting holds---i.e., one where only an inner product structure is assumed---and that a ``structure'' is imposed on it by the requirement that all functions be drawn from some specified class of admissible functions.  Finally it is worth noting that, as defined here the energy norm and Dirichlet integral are equivalent to within a factor of proportionality.  The value of this constant factor is unimportant for minimization problems so long as it is introduced and handled consistently.   

   Given that currently a pre-Hilbert space setting is not generally the first-line choice, its adoption here is worth discussing and justifying at some length.  There are three primary reasons that a structured pre-Hilbert space setting is used here: (1) ease of use issues, (2) non-uniqueness issues and (3) mathematical consistency issues for energy based inner product structures that are defined over unbounded domains.   The ease of use issues can be readily dispatched with:   The main point is that various mathematical derivations can be readily carried out from the given properties of the class of admissible functions.  Along similar lines it should be evident that the pre-Hilbert space setting employed here has certain innate educational benefits.  (In this regard \cite{Mate} gives a fairly complete introduction to RKHS theory based on a pre-Hilbert space setting, but at a higher level than assumed here.)  Each of the remaining two issues will now be addressed in turn.
 
  With regards to the second point, in the real setting is well known that various continuous source distributions inside some bounded region ${\Omega}'$ produce identical fields in the complementary unbounded region $\Omega$ and thus one might expect (\ref{E:TmuA}) to be not only ill-conditioned, but generally ill-posed.  What is less well known is that a proof exists in the geophysical literature that point source basis functions are independent---or, in other words, that a localized finite number of non-zero bounded scalar point sources inside some given fixed sphere of $\mathbb{R}^n$, for $n \geq 2$, cannot exist that produces a null field in the exterior of the same sphere \cite{geoPMuniq}.  (\cite{SIAMuniqArt} discusses a different type of inverse source problem for \R2 logarithmic potentials where the shape of the source region is to be determined.)  Obviously this implies that (\ref{E:TmuA}) is theoretically invertible for point source basis functions.

  A full discussion of point source uniqueness will be given in another venue, but observe that by using (\ref{E:sigLLSQ}) through (\ref{E:TDet}) it is a simple matter to deduce that point pole distributions in both \C\ and \R2 are unique.  In particular, consider the kernel $1/(z - z_k)$ in the complex setting first and suppose, to the contrary, that a pole distribution of the form $\varphi = \sum_{k=1}^{N_k} {{\mu}_k}/({z - z_k})$ exists where $\varphi(z) = 0$ holds for all $z \in \Omega$ (i.e., $|z| \geq 1$) and all the source terms are assumed to be nonzero.  Thus $\|\,\varphi\,\|^2 \eq {\mathbf{\mu}}^{\dag}\mathbf{T}\,\mathbf{\mu} = 0$ must clearly hold for any choice of norm.  Since $\mathbf{T}$ is a (Hermitian) Gram matrix all of its eigenvalues are real and nonnegative.  Let ${\lambda}_k$ denote each of these $N_k$ eigenvalues, then $|\mathbf{T}| = \prod_{k=1}^{N_k}{\lambda}_k \geq 0$ holds for any norm.  Combining this result with the fact that (\ref{E:TDet}) implies $|\mathbf{T}| \neq 0$ holds for the standard integral norm setting, it is obvious that ${\lambda}_k > 0$ and thus that $\mathbf{T}$ is positive definite in this same norm setting.  Then, by either expanding $\mathbf{\mu}$ in terms of unit eigenvectors and substituting the result into ${\mathbf{\mu}}^{\dag}\mathbf{T}\,\mathbf{\mu}\,$, or by diagonalizing $\mathbf{T}$ in the same form, it is apparent that $\|\,\varphi\,\|_{\sigma} = 0 \implies \mathbf{\mu} = 0\,$.  This contradiction proves the desired general assertion in the complex setting.  For the real \R2 setting the uniqueness proof for dipoles is even easier since it follows immediately from the above result when coupled with the facts that ${\text{Re}}\,\{1/(z - z_k)\} = (x - x_k)/\sqrt{(x - x_k)^2 + (y - y_k)^2}$, ${\text{Im}}\,\{1/(z - z_k)\} = (y - y_k)/\sqrt{(x - x_k)^2 + (y - y_k)^2}$ and that both the real and imaginary parts of a complex function are harmonic.

    While it may be true that point source or point pole distributions are unique, ill-posedness considerations clearly remain and as a practical matter these theoretical existence and uniqueness proofs do not in any sense guarantee viable solutions.  Thus theoretical non-uniqueness issues are by-passed here and a utilitarian perspective is adopted.   In this perspective, the minimal requirement is that $\mathbf{T}^{-1}$ be (accurately) numerically computable, which is the assumption made here.  This immediately shifts the perspective from abstract considerations to condition number and other implementation concerns.  As such, condition number issues will be addressed more fully later at the end of this section.  Also this and other  implementation concerns are addressed in Section~\ref{S:examples}.  To see the inherent difficulties here in a Hilbert space setting consider the limit of a sequence of approximating functions given by $\lim_{N_k  \to \infty} \,$ {\raisebox {1pt} {${\varphi}$}}$\vphantom|_{{}_{\!N_k}}$, where \,{\raisebox {1pt} {${\varphi}$}}$\vphantom|_{{}_{\!N_k}}$ is specified by the right hand side of (\ref{E:LLSQbasis}) with a set of fundamental solution basis functions for ${\mathcal{B}}_k$.  Clearly, in this limit it is possible to approach continuous distributions that are known to be indeterminate.  This is one of many troubling sequences whose limits must be admitted from the completeness requirements placed on a Hilbert space and which a pre-Hilbert space setting by-passes. 

   The third point is that various conditions must be enforced to transform a Dirichlet integral into a proper inner-product structure and a structured pre-Hilbert space setting is the natural way to do this.  Historically speaking, the assumption of a structured pre-Hilbert space setting has been very common in dealing with energy norms, especially for unbounded regions \cite{Bergman}.  There are four primary problems to be overcome.  First, the functions to be approximated and the approximating functions must be well behaved on the boundary of $\Omega$.  Second, $\D[F,\,F]$ is inherently indeterminate; for example, in the real setting consider the fact that $\mathbf{\nabla}(F + C) = \mathbf{\nabla}{F}$ for any constant $C$ [where $\mathbf{\nabla}$ denotes the usual $\mathbb{R}^2$ gradients $\eq (\partial\,/\partial\, x,\,\partial\,/\partial\, y)^T$].  This problem is easily solved by imposing the requirement on the class of admissible functions that $\lim_{|z| \to \infty} F(z) = 0$.   Assuming that $F$ and $G$ are sufficiently smooth, the third problem is that $\D[F,\,G]$ many not be bounded.   Assuming that $\mathbf{\nabla}F$ and $\mathbf{\nabla}G$ are bounded, this is easily solved by requiring that $\mathbf{\nabla}F$ and $\mathbf{\nabla}G$ also tail-off fast enough as distances approach infinity.  Fourth, various smoothness issues are present that can easily be by-passed by imposing harmonicity or by requiring analyticity in the complex setting.  As discussed below, since it is fairly easy to solve all of these problems in the complex setting by employing a power series representation, this is the  approach taken here.  (Researchers can simply infer the appropriate minimal requirements, while the student reader who wishes to consider the required properties from a more concise perspective should consult a standard reference such as \cite{AxlerEtAll} and lecturers should see Item 9 of Appendix A, where the motivation for the approach taken here with regards to such matters is presented.)

  Three other observations are relevant.  First, it is also easy to characterize the class of admissible functions by assuming that they are all generated by well behaved density functions.  Second, for bounded domains it is common to employ a Sobolev space instead of a structured pre-Hilbert space to overcome these Dirichlet integral issues, but this entails a basic modification of the inner product structure itself; i.e., in the first Sobolev norm setting a volume integral over $\Omega$ of the product of $F$ and $G$ is added to $\D[F,\,G]$ and it is then required that the resulting inner product evaluations be bounded.  Third, various properties defining the class of admissible functions need not be concisely stated and, in fact, it may be convenient to have a certain amount of redundancy built into them.  (As elsewhere in this article the reader is reminded that generality and conciseness are not to be confused with rigor, as some are prone to do, since rigor does not necessarily require either of these other two conditions.)  The point is that if ease of use (or exposition) is adopted as a primary goal, then one property or another can be more effectively employed to perform some particular derivation so long as it is clear that the functions involved are inside the overall common intersection of subspaces that results when all of the properties concurrently hold (generally each property allows linear superpositions so that a finite span of functions results since linear combinations of functions that satisfy a property, also satisfy the property).  

 Clearly certain types of functions to be fit may fail to fall within the parameters of the given class of allowed admissible functions.  These exceptional cases can generally be handled by applying special attention to them and introducing modifications on a case by case basis as needed.  It is instructive to examine two concrete examples along these lines.  Towards that end, consider the complex setting, which is the main one assumed in the sequel.  Here the properties for the admissible class of functions are specified by (\ref{E:series1}), but it is clear that there are common analytic functions over the region of interest that do not fall within the requirements of this class of admissible functions.  First, although $f(z) = \ln\, z$ is analytic in $\Omega$, it fails to meet either of the criteria specified in (\ref{E:series1}).  In Section~\ref{S:RealPlane} it will be shown that both these difficulties can be overcome by introducing the idea of a paired point.  This paired point formalism physically amounts to considering ``finite extended poles'' only.  As a second example, consider $f(z) = \sin\, z$, which will be analyzed in more detail in Section~\ref{S:examples} and it will be shown there that this function can be modified to produce a suitable function.  This general issue of admissible functions will be dealt with in more detail in Section~\ref{S:examples}.  Here it is worth emphasizing again that in the sequel the class of admissible real functions is obtained from the class of analytic functions (\ref{E:series1}) by simply taking the real part of this series and as discussed elsewhere this correspondence is one-to-one since such admissible real functions (aside from branch-cut related issues for certain cases) have (\ref{E:series1}) as a unique harmonic completion.  This correspondence is essentially an easy way of circumventing the various issues mentioned above in conjunction with Dirichlet integral based inner products.

  As indicated in the introduction and in Appendix~A, it is assumed that all readers have at least some understanding of functional analysis, but for those readers who are not otherwise comfortable with Dirichlet forms or integrals in $\mathbb{R}^2$ a few additional comments are perhaps in order.  First the actual series representation of the class of admissible real valued functions for the exterior of a unit disk, which are denoted here by $F(x,\,y)$, can easily be explicitly determined as follows.  In what follows $x$ and $y$ are Cartesian coordinates and $r$ and $\theta$ are the corresponding polar coordinates.  Each term of (\ref{E:series1}) can be rewritten as 
\begin{equation}\notag
\frac{a_n}{z^n} = \frac{a_n\,e^{-in\theta}}{r^n} = \frac{a_n\cos\,n\theta -  i\,a_n\sin\,n\theta}{r^n}\ ,
\end{equation}
for $n > 0$ and $|z| \geq 1$, with $z \in \mathbb{C}$, $a_n \in \mathbb{C}$ and $n$ an integer.  It thus follows immediately by taking the real part of (\ref{E:series1}) and reidentifying coefficients that
\begin{equation}\label{E:RealFser}
 F \eq {\text{Re}}\, \{\,f(z)\,\} \eq {\ \text{Re}}\, \{\ \sum_{n=1}^{\,\infty} a_nz^{-n}\ \}\,  = \, \sum\limits_{n=1}^{\ \infty}\,\frac{a'_n\sin\,n\theta + b'_n\cos\,n\theta}{r^n}\ ,
\end{equation}
where $a'_n,\, b'_n \in \mathbb{R}$.  Observe that (\ref{E:RealFser}) is just a standard harmonic series representation over the exterior of a unit disk of a harmonic function that vanishes at infinity.  One important property of any admissible $F$ is that it must be bounded with respect to the inner product and norm structure: $\|F\|^2 \eq (F,\,F) < \infty$.  This is accomplished here through the properties of the series forms (\ref{E:RealFser}) and (\ref{E:series1}); hence, some natural but not otherwise obvious constraints must also be placed on the coefficients in these series representations so that any coefficient series that is the result of an inner product or norm evaluation converges.

  Next consider the issue of explicitly showing that $\mathbb{R}^2$ Dirichlet integrals of (\ref{E:RealFser}) are bounded over the exterior region of interest: ${\Omega} \eq \{\ x,\,y\ \mid \ r = \sqrt{x^2 + y^2} \geq 1\ \}$.  Before considering inner products based on Dirichlet integrals consider the following inner product, which will be introduced at the end of Section~\ref{S:SIDACKS} in a different setting:
\begin{equation}
(F,\,G){\lsm}_B = \iint\limits_{{\Omega}}\, F\,G\,dA\ ,
\end{equation}
where $dA$ is the element of area, so that $dA = dx\,dy$ in Cartesian coordinates and  $d\,A = r\,dr\,d\theta$ in standard polar coordinates.  This inner product, however, is unsatisfactory for the class of functions considered here since even the first term in the series on the right hand side of (\ref{E:RealFser}) leads to an unbounded integral:
\begin{equation}\label{E:Divergent}
\|\sin\,\theta/r\|{\ls}_B^2 = \int\limits_{r=1\ }^{\,\,\infty}\int\limits_{\theta=0}^{\,\,2\pi} \,\frac{{\sin}^2\,\theta}{r^2}\ r\,d\,r\,d\,\theta
\end{equation}
(which is clearly unbounded).  Alternatively, since in polar coordinates
\begin{equation}\notag
\mathbf{\nabla} F\ =\ \frac{\partial{F}}{\partial\,r}\,\,\widehat{\text{\begin{Large}{\textbf{e}}\end{Large}}}_{\mathbf{r}}\,\, + \,\, \frac1r\frac{\partial{F}}{\partial\,\theta}\,\,\widehat{\text{\begin{Large}{\textbf{e}}\end{Large}}}_{\mathbf{{\theta}}}\,\ \ \text{and thus}\ \ |\mathbf{\nabla} F|^2\ =\ \Bigg|\frac{\partial{F}}{\partial\,r}\Bigg|^2\,\, + \,\, \frac1{r^2}\,\Bigg|\frac{\partial{F}}{\partial\,\theta}\Bigg|^2\,,
\end{equation}
it is clear that (\ref{E:ReDirInt}) will result in a bounded integral for well behaved $a'_n$ and $b'_n$ so that a consistent norm and inner product structure can be based on (\ref{E:RealFser}) for Dirichlet integrals. 

 To guarantee boundedness of inner product and norm evaluations, it is also necessary to place restrictions on the series coefficients of (\ref{E:RealFser}) and (\ref{E:series1}).  The actual properties that the coefficients $a_n$ in (\ref{E:series1}) or  $a'_n$ and $b'_n$ in (\ref{E:RealFser}) must satisfy can be easily found by direct substitution.  For example, if (\ref{E:RealFser}) is substituted directly into $\D[F,\,F]$ given by (\ref{E:ReDirInt}) then the resulting integrals can be directly evaluated by the same procedures used in Section~\ref{S:SIDACKS}.  This is not done here to emphasize the point that so long as the associated norms are bounded not only are these actual expressions not required here, but that all that is required is that a series of the form (\ref{E:RealFser}) or (\ref{E:series1}) must exist in \underline{principle} so that even the form of the coefficients $a_n$ in (\ref{E:series1}) or of $a'_n$ and $b'_n$ in (\ref{E:RealFser}) need not be known explicitly.

  Finally, from an applied perspective the requirement that solutions to (\ref{E:TmuA}) be numerically computable is clearly significant for both inverse source determination and functional approximation problems, but condition number considerations differ somewhat for these two areas of interest.  In modeling  problems one wishes to minimize the modeling error directly and by introducing a cost function that minimizes this modeling error it can be done effectively; moreover, the minimization of (\ref{E:PHI}) that is implied by (\ref{E:TmuA}) ensures that the modeling match is good regardless of the size of the condition number that is associated with the underlying matrix equation set (at least up to a point).  The underlying issues here can perhaps be most clearly understood by way of an analogy with a standard LLSQ process (i.e., not a generalized LLSQ one), where a Euclidean sum of squares of errors is minimized and the linear system of equations that results is, say $\mathbf{A}\mathbf{x} = \mathbf{b}$.  Here the vector $\mathbf{b}$ is known and the ill-conditioned matrix $\mathbf{A}$ is known, but the vector of parameters $\mathbf{x}$ is unknown and is to be determined.  Let $\mathbf{\tilde{x}}$ denote the true value of $\mathbf{x}$ and $\mathbf{\hat{x}}$ the numerical solution.  Then as an optimization criteria, clearly there is a large difference between requiring that $|\mathbf{A}\mathbf{\hat{x}} - \mathbf{b}|^2$ be small for a successful solution and requiring that $|\mathbf{\hat{x}} - \mathbf{\tilde{x}}|^2$ be small: it is this difference that underlies the point made above.  The underpinning issues here will be addressed further in Section~\ref{S:examples}, as well as in other articles in the sequence.

\section{Standard Integral Norm Setting and General Motivation}\label{S:SIDACKS}

  This section delineates the basic SIDACKS approach.  In what follows it is assumed that functions are bounded and analytic in the exterior and on the boundary of some unit disk specified by $\Omega \eq \{ z \in  \mC\mid\, |z\,| \geq 1 \}$.  It is further assumed that the magnitude of such functions fall off at least as fast as $1/|z|$ as $|z| \to \infty$. (As mentioned in Section~\ref{S:PreDIDACKS} this condition is somewhat restrictive---for example, if $\lim_{|z| \to \infty} f(z) =$ constant, then the constant term can be subtracted off and later added back on after the fit has been performed; however, even after this modification many analytic functions will still fail to meet   this general criteria [see Section~\ref{S:examples} for various examples].)  For $|z| \geq 1$, such functions can be efficiently represented by a power series of the form:
\begin{equation}\label{E:series1}
 f(z) = \sum\limits_{n=1}^{\infty} a_nz^{-n}\ \implies\negmedspace\! \lim_{\,\,\ \ |z| \to \infty}\, f(z) \rightarrow 0 \,,
\end{equation}
where $a_n \in \mC$.  It is assumed that the series in (\ref{E:series1}) is well behaved (i.e., that it yields finite norms) and that it can be operated on termwise to yield series that are bounded and point-wise convergent, which is to say that $\|f\| < \infty$ and that the power series for $F(z) = f(1/z)$ has a radius of convergence that is less than one.     
Equation~(\ref{E:series1}) essentially specifies the admissible class of analytic functions that will be  considered.  It has been stated in a somewhat redundant fashion for emphasis (as discussed in Section~\ref{S:PreDIDACKS} and Item 9 of Appendix A, redundancy here is not necessarily considered undesirable).  In the sequel, although this power series is assumed to exist, none of the actual terms in the series are explicitly required provided that there are no poles located at the origin.  While (\ref{E:series1}) has been written in terms of the function to be approximated, the approximating function is composed of linear combinations of basis functions, which are simple poles with specified positions that clearly satisfy (\ref{E:series1}):
\begin{equation}\label{E:fitform}
 \varphi = \sum\limits_{k=1}^{N_k} \frac{{\mu}_k}{z - z_k}\ .
\end{equation}
Here it is assumed that ${\mu}_k \in \mC$ and that $0 < |z_k| < 1$.  While $z_k \neq 0$ is assumed for convenience, this restrictions can easily be removed (see Section~\ref{S:examples}).  Using the geometric series, each of the pole terms appearing in (\ref{E:fitform}) can be reexpressed as 
\begin{equation}\label{E:geoseries}
\frac{1}{z - z_k} = \sum\limits_{n=0}^{\infty} \frac{z_k^n}{z^{n+1}}\ .
\end{equation}

  As discussed in Section~\ref{S:PreDIDACKS}, the basic strategy is to introduce an appropriate LCF norm structure and then to  minimize $\Phi \eq \|f - \varphi\|^2 = \|f\|^2 - 2(f,\,\varphi) + \|\varphi\|^2$.  Towards that end consider the following inner product:
\begin{equation}\label{E:sigg1}
(f,\,g){\ls}_{\sigma} \eq  \frac1{2\pi} \int\limits_{\theta=0}^{2\pi} [f^*(z)\,g(z)]\Big|_{r=1} d\,\theta\ ,  
\end{equation}
where polar coordinates have been introduced (i.e., $z = re^{i\theta}$, $z_k = r_ke^{{i\theta}_k}$).  [The factor of $1/(2\pi)$ has been inserted here since it is clearly inappropriate to lump it in with the kernel $1/(z -z_k)$---as is the usual convention in RKHS theory---and the alternative is to allow factors of $\pi$ in the point replication property itself.]  Clearly when $\| f - \varphi \|_{\sigma}^2 \eq (f - \varphi,\,f - \varphi)_{\sigma}^2 = (\,|f - \varphi|\,,\,|f - \varphi|\,)_{\sigma}^2\, \eq\ \|\,\,|f - \varphi|\,\,\|_{\sigma}^2 $ is minimized the boundary conditions on the unit disk are also minimized and thus $\varphi \approx f$ follows immediately from the uniqueness of Dirichlet boundary conditions.  Replacing $g$ by $1/(z - z_k)$ in (\ref{E:sigg1}) and using (\ref{E:series1}) and (\ref{E:geoseries}) yields:
\begin{equation}\label{E:sigg2}
(f,\,(z - z_k)^{-1}){\ls}_{\sigma} = \frac1{2\pi}\int\limits_{\theta=0}^{2\pi}\sum\limits_{n=1}^{\infty}\,a^{*}_n\,e^{i\,n\theta}\ \sum\limits_{n'=0}^{\infty}\,z_k^{n'}e^{-i(n'+1)\theta}\ d\,\theta
= \frac1{z_k}\sum\limits_{n=1}^{\infty}\,a^{*}_n z_k^{n}\ .
\end{equation}
Introducing $p_k = 1/z_k$ allows (\ref{E:sigg2}) to be rewritten in the final form
\begin{equation}\label{E:sigg3}
 (f,\,(z - z_k)^{-1}){\ls}_{\sigma} = p_k\,(f(p^{*}_k))^{*}\ .
\end{equation}

 When (\ref{E:fitform}) is used to reexpress $\varphi$ and the result is substituted into ${\Phi}_{\sigma} \eq \|f - \varphi\|_{\sigma}^2$, a bi-linear form in terms of ${{\mu}_k}^{*}$ and ${\mu}_k$ results, which can be further simplified by introducing $((z - z_k)^{-1},\,f)_{\sigma} \eq A_k$ and $((z - z_{k'})^{-1},\,(z - z_k)^{-1})_{\sigma} \eq T_{k'\!,\,k}$.  The resulting expression for ${\Phi}_{\sigma}$ can be further simplified since $A_k$ and $T_{k'\!,\,k}$ can be rewritten as closed form expressions by using (\ref{E:sigg3}).  [Here $T_{k'\!,\,k} = (p_kp^{*}_{k'})/(p_kp^{*}_{k'} - 1) = (1 - z_kz^{*}_{k'})^{-1}$ and $A_k = p^{*}_kf(p^{*}_k)$.]  Thus it is only necessary to take partials of this bi-linear form with respect to the source parameters and then set the result to zero in order to obtain a closed-form equation set which gives the values of the pole strengths (${\mu}_j$) that minimize  ${\Phi}_{\sigma}$.  Specifically, if $\mathbf{T}$ denotes the matrix whose elements are $T_{k'\!,\,k}$, $\mathbf{\mu}$ the vector whose elements are ${\mu}_k$ and $\mathbf{A}$ the vector whose element are $A_k$ then, as shown in Section~\ref{S:DIDACKS} using complex conjugate variables, the linear equation set governing the source strengths can be written as
\begin{equation}\label{E:sigLLSQ}
{\mathbf{T}}_{\smallindex{[\sigma]}}\,{\mathbf{\mu}}_{\smallindex{[\sigma]}}\, = {\mathbf{A}}_{\smallindex{[\sigma]}}.
\end{equation}
In (\ref{E:sigLLSQ}) a generally omitted [optional] $\sigma$ subscript is indicated: i.e., ${\mathbf{T}}_{\smallindex{[\sigma]}} \eq {\mathbf{T}}$,\,\, ${\mathbf{\mu}}_{\smallindex{[\sigma]}} \eq {\mathbf{\mu}}$\, and\, ${\mathbf{A}}_{\smallindex{[\sigma]}} \eq \mathbf{A}$.

 Equation~(\ref{E:sigLLSQ}) can also be obtained without resorting to complex conjugate coordinates by following a more involved sequence of steps.  First introduce variables for the real and imaginary components of ${\mu}_k$: ${\alpha}_k \eq {\text{Re}}\,\{{\mu}_k\}$ and ${\beta}_k \eq {\text{Im}}\,\{{\mu}_k\}$.  This yields a cost function ${\Phi}_{\sigma} = {\Phi}_{\sigma}({\alpha}_j,\,{\beta}_j)$ with real arguments.  Next take partials with respect to ${\alpha}_j$ and ${\beta}_j$ and set the results to zero, which yields two real coupled linear equation sets that can be rewritten as a single real equation set in block matrix form:
\begin{equation}\label{E:realT}
\left(
\begin{array}{r|l}
 {\mathbf{T}}_{\text{R}_{\vphantom{[}}} & {\mathbf{T}}_{\text{I}}\\ \hline
 -{\mathbf{T}}_{\text{I}}^{\vphantom{[}} & {\mathbf{T}}_{\text{R}}
\end{array}  
\right)\!
\left(
\begin{array}{c}
 {\mathbf{\alpha}}_{{\vphantom{R}}_{\vphantom{[}}}\\ \hline
 \mathbf{\beta}_{\vphantom{R}}^{\vphantom{[}}
\end{array}
\right)\,
 = 
\left(
\begin{array}{c}
 {\mathbf{A}}_{{\text{R}}_{\vphantom{[}}}\\ \hline
 {\mathbf{A}}_{\text{I}}^{\vphantom{[}}
\end{array}
\right)\ .
\end{equation}
In (\ref{E:realT}) the subscript R on a vector or matrix denotes the real part of the vector or matrix in question, and I denotes the imaginary part.  Also ${\mathbf{\alpha}}$ denotes the $N_k$ vector whose components are ${\alpha}_k$ and ${\mathbf{\beta}}$ the $N_k$ vector whose components are ${\beta}_k$.
Finally, by substituting ${\mathbf{T}} = {\mathbf{T}}_{\text{R}} + i\,{\mathbf{T}}_{\text{I}}$,
$\mathbf{\mu} =  {\mathbf{\alpha}} + i\,{\mathbf{\beta}}$ and ${\mathbf{A}} = {\mathbf{A}}_{\text{A}} + i\,{\mathbf{A}}_{\text{I}}$ into (\ref{E:sigLLSQ}) it is a trivial matter to show it is equivalent to (\ref{E:realT}) \cite{NumericalRecipes}.

Here $\mathbf{T}$ can be rewritten as the product of a matrix $\mathbf{L}$ whose elements are $L_{k',\,k} = 1/(p_k - z_{k'}^{*})$ and a diagonal matrix, $\mathbf{D_p}$ whose elements are ${\delta}_{i,\,j}\,p_j$:
\begin{equation}
 \mathbf{T} = \mathbf{L}\mathbf{D_p}\ .
\end{equation}
The determinant of $\mathbf{L}$, $|\mathbf{L}|$, is the well-known Cauchy determinant.  Introducing the standard symbols  $a_i \eq p_i$ and $b_j \eq  -z_{j}^{*}$ used in    closed-form expressions of Cauchy's determinant, the associated matrix has elements of the form $1/(a_i + b_j)$ and the closed-form expressions itself can be written as \cite[p. 268]{PJDavis}
\begin{equation}\label{E:CauchyDet}
 |\mathbf{L}| = \frac{\prod\limits_{i>j}^{N_k} (a_i - a_j)(b_i - b_j)}{\prod\limits_{i,\,j=1}^{N_k}(a_i + b_j)} = \frac{\prod\limits_{i>j}^{N_k} (p_i - p_j)(z_{j}^{*} - z_{i}^{*})}{\prod\limits_{i,\,j=1}^{N_k}(p_i - z_{j}^{*})}\ .
\end{equation}
While (\ref{E:CauchyDet}) is frequently stated for $a_i$ and $b_j \in \mRR$,  since the standard proof is by induction and involves normal algebraic manipulation \cite{Deutsch} it clearly holds for $a_i$ and $b_j \in \mC$ as well.  Combining $|\mathbf{D_p}| = p_1p_2p_3\cdots p_{N_k}$ with this closed-form expression gives: 
\begin{equation}\label{E:TDet}
 |\mathbf{T}|  =  \frac{\prod\limits_{i>j}^{N_k} (p_i - p_j)(z_{j}^{*} - z_{i}^{*})}{\prod\limits_{i,\,j=1}^{N_k}(p_i - z_{j}^{*})}\,\,\prod\limits_{k=1}^{N_k}p_k\ .
\end{equation}
 Clearly $|\mathbf{T}| \neq 0$ when $0 < |z_k| < 1$ since $z_k \neq z_{k'}$ holds for $k' \neq k$.   Thus ${\mathbf{T}}^{-1}$ exists, but as discussed in Sections~\ref{S:PreDIDACKS} and \ref{S:examples} some care is normally required in solving (\ref{E:sigLLSQ}) or (\ref{E:realT}).

Higher order poles can be easily incorporated into the fitting function $\varphi$ since
\begin{equation}\label{E:HOP}
 \frac{d^m\ }{d\,z^m} \bigg(\frac1{z - z_k}\bigg) = \frac{(-1)^m\,m!}{(z - z_k)^{m+1}} = -\frac{d^m\ }{d\,z_k^m} \bigg(\frac1{z - z_k}\bigg)\ 
\end{equation}
and when partials with respect to $z_k$ are taken on both sides of (\ref{E:sigg3}) analogous closed-form expressions are produced for higher order poles.  For example, a general fitting function of the form 
\begin{equation}\label{E:fitform2}
 \varphi = \sum\limits_{k=1}^{N_k}\ \sum\limits_{m=1}^{N_m(k)}\frac{{\nu}_{k,\,m}}{(z - z_k)^m}
\end{equation}
can be easily handled, where the ${\nu}_{k,\,m}$'s are the different pole strengths of various orders at distinct locations.  (Here ${N_m(k)}$ is the number of different types of poles at location $z_k$.)  Clearly when this form is substituted in place of (\ref{E:fitform}) into ${\Phi}_{\sigma}\eq \|f - \varphi\|_{\sigma}^2$ a set of closed-form equations result for these various higher order pole strengths.

  So far the development has focused on the exterior of a unit disk; nevertheless, from an applications perspective a number of generalizations are possible.  For example, if the region of interest consists of the exterior of a disk of different size that is centered over a different point then obviously a translation and rescaling can be applied beforehand and then, after a fit has been performed, the results can be rescaled and translated back to the original coordinates.  Another fairly obvious approach is to employ (\ref{E:sigLLSQ}) to implement either a local or global interpolation scheme.  Such schemes can be based on the fitting function forms given by (\ref{E:fitform}) or more general ones such as (\ref{E:fitform2}).  Obviously the specified data for such interpolation schemes---including the function itself and its derivatives up through order $[{N_m(k)} - 1]$---is matched at all of the conjugate involution points. 
   
  As mentioned in Section~\ref{S:intro}, the basic SIDACKS relationships---especially (\ref{E:sigg3})---are disarmingly simple and it is natural to wonder if and how they are related to known results.  The most direct and obvious connection is of (\ref{E:sigg3}) to the Cauchy integral formula.  The Cauchy integral formula for the interior of a unit disk can be written as 
\begin{equation}\label{E:Cauchy}
F(\zeta) = \frac1{2\pi i}\negmedspace \oint\limits_{\vphantom{\big]}|w|=1}\!\!\frac{F(w)}{w - \zeta}\,\,d\,w\ ,
\end{equation}
where $|\zeta| < 1$ and $F(\zeta)$ is analytic.  Introducing $z = 1/{\zeta}$ and relabeling $F({\zeta}) \eq F(1/z) \eq f(z)$ in (\ref{E:Cauchy}) yields (\ref{E:sigg3}).  Several RKHSs for either the interior or exterior region of the unit disk can also be tied to Cauchy's integral formula so the SIDACKS approach is either directly or indirectly related to them as well.   Also as previously mentioned, the most significant RKHS connections are to the Bergman kernels for both the interior and exterior of a unit disk and to Szeg\H{o} kernel theory.  Since Bergman kernels for \R2 are analogous to those for \C, this later case is focused on in the sequel.  These RKHS structures, of course, have RCF inner products and kernels.   

 The link between the Bergman kernel for the interior of a unit disk ($K_B$) is worth briefly considering since it too has links to the Cauchy integral formula.   Let ${\bar{\Omega}}' \eq {\Omega}' \cup \partial {\Omega}'$, then the Bergman inner product for the unit disk (${\Omega}'$) is commonly written in RCF and is
\begin{equation}\label{E:BergmanInt}
(F,\,G){\ls}_{B} \eq \iint\limits_{{\bar{\Omega}}'}F(w)\,(G(w))^{*}\,\,d\,A
\end{equation}
where the area integration is performed over the components of $w$ with $w \in {\bar{\Omega}}'$.  The Bergman kernel is then
\begin{equation}\label{E:BergmanKern}
 K_B(\zeta,\,w) = \frac1{\pi}\frac1{(1 - {\zeta}^{*}w)^2}\,,
\end{equation}
where, as above, $\zeta < 1$.  As will be shown next, this kernel has the reproducing property
\begin{equation}\label{E:BKrep}
(F,\,K_B){\ls}_{B} = F(\zeta)\,.
\end{equation}
First observe that the derivative of the geometric series $1/(1 -x) = \sum_{n=0}^{\infty} x^n$ yields 
\begin{equation}\notag
\frac1{(1 -x)^2} = \sum_{n=0}^{\infty} (n + 1)\,x^n\,,
\end{equation}
 so (\ref{E:BergmanKern}) can be rewritten as 
\begin{equation}\label{E:BKseries}
 K_B(\zeta,\,w) = \frac1{\pi}\,\sum\limits_{n=0}^{\infty} ({\zeta}^{*}w)^n\ .
\end{equation}
From (\ref{E:BKseries}) one can easily show
\begin{equation}\label{E:Zpower}
{\zeta}^n = \iint\limits_{{\bar{\Omega}}'}w^n\,K_B^{*}\,\,d\,A
\end{equation}
by rewriting  $d\,A$ in terms of polar coordinates and then performing the indicated integration over the unit disk, as was done, for example, in deriving (\ref{E:sigg2}).
Next, let $\eta > 1$, then substituting the geometric series for $1/(\eta - w)$ into $(\,\cdot\,,\,K_B){\ls}_B$ and using (\ref{E:Zpower}), it follows immediately that
\begin{equation}\label{E:CauchyDenom}
(\,(\eta - w)^{-1},\,K_B){\ls}_B = \frac1{(\eta - \zeta)}\ .
\end{equation}
To show the connection to the Cauchy integral formula consider a closed curve $\gamma$ which is completely outside the unit disk and suppose that $F(z)$ is analytic inside and on $\gamma$, then the Cauchy integral formula for a point $w$ which is inside the unit disk can be written as
\begin{equation}\label{E:CI2}
F(w)\, = \, \frac1{2\pi\,i} \oint\limits_{\gamma} \frac{F(\eta)}{\eta - w}\,\,d\,\eta \ .
\end{equation}
From (\ref{E:CauchyDenom}) and (\ref{E:CI2}) it follows immediately that
\begin{equation}\label{E:FBK}
(F,\,K_B){\ls}_B = \, \frac1{2\pi\,i} \oint\limits_{\gamma} {F(\eta)}\,\Big(\,({\eta - w})^{-1},\,K_B\Big)_B\,\,d\,\eta \, = \,\frac1{2\pi\,i} \oint\limits_{\gamma} \frac{F(\eta)}{\eta - \zeta}\,\,d\,\eta = F(\zeta)\,,
\end{equation}
which is the desired reproducing property.  Detailed rigorous discussions of Bergman kernel theory can be found in various entry level books  \cite{Conformal,PJDavis}.
 It is also worth noting that although Bergman spaces, which are an outgrowth of basic Bergman kernel theory, have been around for more than fifty years \cite{BergmanKern}, they have continued to be an area of intermittent research that has fairly recently resurged \cite{BergmanSpaces2,BergmanSpaces}. 

  Just as was done with (\ref{E:FBK}) for disk interiors, the Bergman kernel for the area over the exterior region of a unit disk can also be derived from the Cauchy integral formula.  Thus the DACK for the standard norm and the Bergman kernel for the interior of a unit disk are linked through the Cauchy integral formula itself. It is also true that the Dirichlet integral based energy norm kernels, which are introduced in the sequel, can also be linked to the Cauchy integral formula, but in this case a more direct connection exists to the Bergman kernel for the exterior of a unit disk and this connection will be explored at the end of Section~\ref{S:DIDACKS}.

  Links also exist between the standard DACKS formalism, Hardy space (${\text{H}}^2$) and Szeg\H{o} kernels.  When equipped with the Szeg\H{o} kernel, ${\text{H}}^2$ can be considered a RKHS.  In particular, the Szeg\H{o} kernel inserted into the standard Hardy inner product [\,denoted $(\,\,\cdot\,\,,\,\,\cdot\,\,)_{\text{H}}^2$\,] yields an alternative form of the Cauchy integral formula; however, an even closer connection exists between the pre-Hilbert form of ${\text{H}}^2$ and the standard DACKS formalism.  This connection is worth considering in more  detail since the Cauchy integral formula need not be employed as an intermediary.

  First, in order to fix the notation it is useful to explicitly introduce the Szeg\H{o} kernel and its associated inner product spaces.  Following M\'{a}t\'{e} \cite{Mate}, let ${\text{H}}_0^2$ denote the pre-Hilbert space of functions analytic in the interior of a unit disk ($|z| < 1$) and continuous on the boundary $|z| = 1$, with inner product specified by
\begin{equation}\label{E:HoTwo}
(f,\,g){\ls}_{{\text{H}}_0^2}\, \eq \,  \frac1{2\pi\,i}\negmedspace \oint\limits_{\vphantom{\big]}|z|=1}\!\! {f(z)\,g^*(z)}\,\,(1/z)\ dz \ .
\end{equation}
Here this line integral expression for the ${\text{H}}_0^2$ inner product has been written in standard RCF to facilitate cross referencing.  Since $|z| = 1$ is understood in the integrand occuring in (\ref{E:HoTwo}), the relation $dz = i\,e^{i\,\theta}\, d\,\theta = i\,z\,\,d\,\theta$  can be used to yield the equivalent form  
\begin{equation}\label{E:Ltwo}
(f,\,g){\ls}_{{\text{H}}_0^2} =  \frac1{2\pi} \int\limits_{\theta=0}^{2\pi} {f(z)\,g^*(z)}\,d\,\theta \ ,
\end{equation}
which is the standard form of the inner product for the pre-Hilbert space $L_0^2$ over $[0,\,2\pi]$, with $f(2\pi) = f(0)$ understood.  The completion of $L_0^2$ is, of course, the Hilbert space $L^2$.  Likewise the completion of ${\text{H}}_0^2$ is denoted  ${\text{H}}^2$ and it is an RKHS (given natural restrictions).   As just noted the reproducing kernel for this RKHS is the Szeg\H{o} kernel $K_{S}$.  Thus substituting $g = K_{S} \eq 1/(1 - z\,s^*)$ where $s \in \mC$, with $|s| < 1$, into (\ref{E:Ltwo}) produces
\begin{equation}\label{E:SZrep}
(f,\,K_{S}){\ls}_{{\text{H}}_0^2} =  f(s)\ .
\end{equation}

  The connections to standard SIDACKS theory are easily derived.  While it is understood that SIDACKS theory holds for the exterior of the unit disk instead of the interior, this fact fails to show up explicitly in the inner products themselves (where $|z| = 1$ is understood), unless $f$ or $g$ depends on auxiliary parameters.  Thus, as explained in the following, (\ref{E:Ltwo}) is formally equivalent to the standard SIDACKS inner product given by (\ref{E:sigg1}), except for the use of RCF versus LCF.  There are some subtleties here that involve the definition of admissible functions; however, these issues can be easily clarified after the fact.  Equation~(\ref{E:SZrep}) explicitly involves an auxiliary parameter ($s$) so some care is required.  The procedure will be to recast (\ref{E:sigg3}) from interior LCF into exterior RCF.  The conversion to RCF simply entails taking the conjugate of the expression that results from combining (\ref{E:sigg1}) and (\ref{E:sigg3}):
\begin{equation}\label{E:Csig}
\frac1{2\pi} \int\limits_{\theta=0}^{2\pi} \Big[\frac{f(z)}{z^* - z^{*}_{k}}\Big]\bigg|_{|z|=1}\,\,d\,\theta = p^{*}_k\,f(p^{*}_k)\ .
\end{equation}
For clarity let $z' \eq 1/z$, $s = z^{*}_k$ and $w \eq z' = r'e^{i{\theta}'} = (1/r)e^{ -i\theta}$ so that ({\ref{E:Csig}) can be rewritten as
\begin{equation}\label{E:Csig2}
\frac1{2\pi} \int\limits_{{\theta}'=0}^{2\pi} \Big[\frac{w^*f(1/z'))}{(1 - w^*s)}\Big]\bigg|_{|w|=1}\,\,d\,{\theta}' = f(1/s)/s\ .
\end{equation}
The function $F(w) = f(1/w)/w$ occuring on the right hand side of (\ref{E:Csig2}) is the relevant one and it is thus useful to consider the form of its power series in terms of $w$ [which is based on (\ref{E:series1})]:
 \begin{equation}\label{E:Fseries}
 F(w) = \sum\limits_{n=0}^{\infty} a_{n+1}w^{n} \ .
\end{equation}
Since $w^{*} = 1/w$ for $|w| = 1$, (\ref{E:Csig2}) can be rewritten as
\begin{equation}\label{E:Csig3}
\frac1{2\pi} \int\limits_{{\theta}'=0}^{2\pi} \Big[\frac{F(w)}{(1 - w^{*}s)}\Big]\bigg|_{|w|=1}\,\,d\,{\theta}' = F(s)\ 
\end{equation}
and this obviously corresponds to (\ref{E:SZrep}). The subtlety indicated above in identifying the functions that occur in (\ref{E:sigg1}) and (\ref{E:Ltwo}) is thus clearly tied to the fact that $f$ must actually be replaced by $F$.  Notice that since $|z| = 1$ is understood in the integrand of (\ref{E:sigg1}), a factor of $z^{*}z$ can be inserted to aid in this reidentification process [the converse identification can easily be carried out using (\ref{E:SZrep})]. 

  Having shown a correspondence between inner product and replication kernel structures, the comparison between exterior SIDACKS theory and Szeg\H{o} kernel structures is now complete. 
 
   For fits that do not involve derivatives (i.e., higher order poles) there is no real significant formal difference between SIDACKS fits using the simple-pole form $1/(z_k - z)$ and Szeg\H{o} kernel based interpolation.  Thus for the interior of a unit disk it might be reasonable to consider the kernel $1/(z_k - z)$ as just the SIDACKS form of the Szeg\H{o} kernel; however, this nomenclature would be ambiguous here for various reasons that include labeling inconsistencies of norms and kernels---for example, at the end of Section~\ref{S:IntDACKS} it will be shown that a different norm can be associated with a very similar kernel ($1/[z_k - z] - 1/z_k$) over the same domain studied for $1/(z_k - z)$.  The advantages of RKHS based approaches are well known, but even for interior problems DACKS theory is worth considering in its own right as an alternative to other interpolation schemes since it allows access to a number of diverse results in a easy, natural and seamless fashion.  For example, for the SIDACKS interior case it is easy to use (\ref{E:HOP}) to produce higher-order pole fits, but the same thing is cumbersome using the Szeg\H{o} or Bergman kernel.  
  Also, as discussed in Section~\ref{S:intro}, by providing a different perspective, consideration of aspects that might not otherwise be obvious can be entertained.  Also as implied there, only a small portion of the overall DACKS formalism is contained in this article and much of the motivation for DACKS theory is contained in these other developments. Additionally, DACKS theory has a natural relationship to both energy minimization based approaches and to logarithmic source harmonic problems, which is generally missing from Bergman and Szeg\H{o} kernel theory since potential theory, inverse source theory and Dirichlet integral applications are not normally considered to be a part of Bergman or Szeg\H{o} kernel theory.

\section{DIDACKS Development}\label{S:DIDACKS}

   Let $G(x,y)$ and $H(x,y)$ be two real valued harmonic functions in the exterior of a unit disk in \R2, then the standard real Dirichlet integral for this region is defined as
\begin{equation}\label{E:ReDirInt}
\D[G,\,H] \eq \int\limits_{r=1\ }^{\,\,\infty}\int\limits_{\theta=0}^{\,\,2\pi} \mathbf{\nabla} G\cdot\mathbf{\nabla} H\ \,r\,d\,r\,d\,\theta\ .
\end{equation}
As discussed in Section~\ref{S:PreDIDACKS}, it is expedient to restrict $G$ and $H$ to be in some admissible class of functions so that $\D[G,\,H]$ can be associated with a positive definite inner product.  Since the overall strategy employed here will be to recast (\ref{E:ReDirInt}) into a complex form that is more mathematically amenable, this class of admissible functions will be linked to this complex form.  Thus the class of admissible functions given by (\ref{E:series1}) will be used to specify the functions of interest for this section and these functions are associated with the complex analog of (\ref{E:ReDirInt}) and its various ramifications.  After the developments of this section are complete, the real \R2 setting will be studied in the next section.  The admissible class of \R2 functions introduced there will then be defined by the requirement that $\D[G,\,H]$ and its complex analog be linked through the act of standard completion of $G$ and $H$ (c.f., Definition~\ref{S:RealPlane}.1}.).  The Dirichlet norm for the real setting is labeled the energy norm and it will not be actually
introduced until late in the next section by (\ref{E:energy}).

  Thus consider the following known but underutilized complex generalization of $\D[G,\,H]$ over $\Omega \subset \mC $:
\begin{equation}\label{E:CplxDirInt}
\D_c[f,\,g] \eq \int\limits_{r=1\ }^{\,\,\infty}\int\limits_{\theta=0}^{\,2\pi} \bigg(\frac{d\,f}{d\,z}\bigg)^{*}\bigg(\frac{d\,g}{d\,z}\bigg)\, r\,d\,r\,d\,\theta\ .
\end{equation}
where $f$ and $g$ are analytic in $\Omega$. Let $\delta\! f \eq f - \, \varphi = u_{\delta}(x,\,y) + i\,v_{\delta}(x,\,y)$ define the modeling difference that is to be minimized, where $u_{\delta}$ and $\,v_{\delta}$ are real valued functions of $x$ and $y$.  Since $\delta\! f$ is analytic, $u_{\delta}$ and $\,v_{\delta}$ satisfy the Cauchy-Riemann conditions and so
\begin{equation}\label{E:CmplxReal}
\bigg(\frac{d\,\,\delta \!f}{d\,z\ }\bigg)^{*}\bigg(\frac{d\,\,\delta\! f}{d\,z\ }\bigg) = \bigg(\frac{\partial u_{\delta}}{\partial x}\bigg)^2 + \bigg(\frac{\partial u_{\delta}}{\partial y}\bigg)^2
= \bigg(\frac{\partial v_{\delta}}{\partial x}\bigg)^2 + \bigg(\frac{\partial v_{\delta}}{\partial y}\bigg)^2\ . 
\end{equation}
Thus minimizing ${\Phi}_D \eq \|\,\delta\! f\,\|_{D}^2$ is equivalent to minimizing $\D[u,\,u]$ or $\D[v,\,v]$ , where 
\begin{equation}\label{E:Dnorm}
\|\,\delta\! f\,\|{\ls}_{D} \eq \sqrt{\D_c[\delta\! f,\,\delta\! f]\,/{2\pi}}\ .
\end{equation}
  The associated inner product is, of course, given by $(f,\,g)_D \eq D_c[f,\,g]/(2\pi)$.
(Notice that ${\Phi}_D$ is real valued.)  The factor of $1/({2\pi})$ here has been inserted for convenience and to help facilitate the identification of $\|\,\delta\! f\,\|{\ls}_{D}^2$ with field energy [which, aside from a units dependent constant, generally has a density of $(\mathbf{\nabla} H\cdot\mathbf{\nabla} H)/({2\pi})$].

  Taking the derivatives of (\ref{E:series1}) and (\ref{E:geoseries}) and substituting the result into the right hand side of (\ref{E:CplxDirInt}) yields
\begin{equation}
(f,\,(z - z_k)^{-1}){\ls}_{D} = \frac1{2\pi}\sum\limits_{m=0}^{\infty}\sum\limits_{n=0}^{\infty}
\ \int\limits_{r=1\ }^{\,\,\infty}\int\limits_{\theta=0}^{\,2\pi} \bigg(\frac{(m + 1)\,a^{*}_{m+1}e^{i(m+2)\theta}}{r^{m+2}}\bigg)\bigg(\frac{(n+1)z_k^{n}e^{-i(n+2)\theta}}{r^{n+2}}\bigg)\ \,r\,d\,r\,d\,\theta
\end{equation}
where $m$ was reindexed to start at $0$ rather than $1$.  Clearly
\begin{equation}\label{E:CollPropC}
(f,\,(z - z_k)^{-1}){\ls}_{D} = \sum\limits_{n=0}^{\infty}{(n+1)^2a^{*}_{n+1}z_k^n}\,\,\int\limits_{r=1\ }^{\,\,\infty}\frac{r}{(r^{n+2})^2}\,d\,r = \frac1{2z_k}\sum\limits_{n=1}^{\infty}n\,a^{*}_{n}z_k^n = -\frac1{2z_k}\bigg(\Blbrac\frac{d\,f}{d\,z}\bigg]\bigg|_{z=1/z_k^{*}}\bigg)^{*}
\end{equation}
Introducing $p_k \eq 1/z_k$ and $f_z(z) \eq {d\,f}/{d\,z}(z)$, (\ref{E:CollPropC}) can be reexpressed as
\begin{equation}\label{E:CollPropD}
(f,\,(z - z_k)^{-1}){\ls}_{D} = -\frac{p_k^2}2\,\{f_z(p_k^{*})\}^{*} \eq -\frac{p_k^2}2\,f_z^{*}(p_k^{*})\ ,
\end{equation}
which is analogous to (\ref{E:sigg3}).  As before, let $A_k \eq ((z - z_k)^{-1},\,f)_{D} \eq (f,\,(z - z_k)^{-1})_{D}^{*}$ and $T_{k',\,k} \eq ((z - z_{k'})^{-1},\,(z - z_{k})^{-1}){\ls}_D\,$, then $\Phi_D$ can be written as
\begin{equation}\label{E:p}
\Phi_D = \|\,f\,\|{\ls}_D^2 + \sum\limits_{k=1}^{N_k}\sum\limits_{k'=1}^{N_k} {\mu}_k^{*}{\mu}_{k'}\,T_{k,\,k'} - \frac12\sum\limits_{k=1}{\mu}_kA^{*}_k - \frac12\sum\limits_{k=1}{\mu}_k^{*}A_k\ .
\end{equation}
Equation~(\ref{E:CollPropD}) can be used to obtain explicit expressions for $A_k$ and  $T_{k',\,k}$: 
\begin{equation}
A_k  \eq - (p^{*}_k)^2\,f_z(p_k^{*})\,\ \ {\text{and}}\ \ 
T_{k',\,k} = \frac12 \frac{(p_kp_{k'}^{*})^2}{(p_kp_{k'}^{*} - 1)^2}\ .
\end{equation}

  Introducing the matrix $\mathbf{T}$ and vectors $\mathbf{A}$ and $\mathbf{\mu}$, as in (\ref{E:TmuA}) or (\ref{E:sigLLSQ}), along with the pair of complex conjugate variables ${\mu}_k$ and  $\bar{\mu}_k \eq {{\mu}_k}^{*}$ into (\ref{E:p}) yields a cost function of the form ${\Phi}_D = {\Phi}_D(\bar{\mathbf{\mu}},\,\mathbf{\mu})$.  The minimum of ${\Phi}_D$ is then found by setting $\partial\,{\Phi}_D/\partial\, \bar{\mu}_j = 0$ (for $j = 1,\,2,\,3\,\cdots,\,N_k$):
 \begin{equation}\label{E:DLLSQ}
{\mathbf{T}}_{\smallindex{[D]}}\,{\mathbf{\mu}}_{\smallindex{[D]}} = {\mathbf{A}}_{\smallindex{[D]}}, 
\end{equation}
in accord with (\ref{E:sigLLSQ}).  
In (\ref{E:DLLSQ}) an [optional] $D$ subscript is shown which is generally omitted: ${\mathbf{T}}_{\smallindex{[D]}} \eq {\mathbf{T}}$, ${\mathbf{\mu}}_{\smallindex{[D]}} \eq {\mathbf{\mu}}$ and ${\mathbf{A}}_{\smallindex{[D]}} \eq \mathbf{A}$.
Obviously, applications here follow along the same lines as those of (\ref{E:sigLLSQ}) so that, for example, fits based on (\ref{E:HOP}) and (\ref{E:fitform2}) can be carried out here in the same fashion.
  
  For any suitable domain in the complex setting, when $\text{D}_c[f,\,g]$ is defined analogously to (\ref{E:CplxDirInt}) an intimate connection exists between $\text{D}_c[f,\,g]$ and Bergman kernel theory for the same domain.  For example, in analogy to (\ref{E:BergmanInt}), when $f$ and $g$ are in the admissible class of functions, consider the following RCF Bergman inner product for the exterior of a unit disk:  
\begin{equation}
(f,\,g){\ls}_{\!B} \eq \int\limits_{r=1\ }^{\,\,\infty}\int\limits_{\theta=0}^{\,\,2\pi} f(z)(g(z))^{*} r\,dr\,d \theta\ .
\end{equation}
Since the derivatives of $f$ and $g$ always exist, introduce $F(z) \eq {d\,f}/{d\,z}\,$ and $G(z) \eq {d\,g}/{d\,z}\,$. 
Then with $p_k \eq 1/z_k$ and $f_z(z) \eq {d\,f}/{d\,z}\,$ this inner product yields
\begin{equation}\label{E:DirichletBergman1}
(F,\,G){\ls}_{\!B}^{*}/2\pi =
(f,\,g){\ls}_{D}\,\,, 
\end{equation}
which clearly holds in general for any domain in the complex plane (with suitable restrictions on the class of admissible functions).

\section{Laplacian Inverse Source Theory in the Real Plane}\label{S:RealPlane}

   This section transcribes results obtained in Sections \ref{S:SIDACKS} and \ref{S:DIDACKS} for the complex setting to the real \R2 setting.  A direct \cite{SIAMuniqArt}, but slightly nonstandard, way to map a harmonic function specified over some region of \R2, say $W(x,\,y)$, into an analytic function $f_{\smallindex{W}}$ that is defined over the corresponding region of \C\ is to set $f_{\smallindex{W}}(z) \eq u_{\smallindex{W}}(x,\,y) + i\, v_{\smallindex{W}}(x,\,y)$, for real valued functions $u_{\smallindex{W}}$ and $v_{\smallindex{W}}$, where
\begin{equation}\label{E:mappA}
 u_{\smallindex{W}}(x,\,y) \eq \frac{\partial W}{\partial y}\ \ \ \text{and}\ \ \ v_{\smallindex{W}}(x,\,y) \eq \frac{\partial W}{\partial x}\ .
\end{equation}
Using this correspondence in (\ref{E:sigg1}) immediately yields
\begin{equation}\label{E:sigg5}
\|f_{\smallindex{W}}\|^2_{\sigma} =  \frac1{2\pi} \int\limits_{\theta=0}^{2\pi} \Blbrac\bigg(\frac{\partial W}{\partial x}\bigg)^2 + \bigg(\frac{\partial W}{\partial y}\bigg)^2\bigg]\Bigg|_{r=1} d\,\theta\ ,  
\end{equation}
so that minimizing ${\Phi}_{\sigma} =\|f_{\smallindex{W}}\|^2_{\sigma}$ enforces Neumann boundary conditions on $W$.  Note, however, that minimizing $\|f_{\smallindex{W}}\|_D^2$ implies minimizing $\D[{\partial W}\!/{\partial x},\,{\partial W}\!/{\partial x}]$ or $\D[{\partial W}\!/{\partial y},\,{\partial W}\!/{\partial y}]$ rather than $\D[W,\,W]$ itself, so that the analytic correspondence given by (\ref{E:mappA}) will not be used in the sequel. 

  As an alternative, thus consider the standard way to map $W$ into an analytic function:

\vskip 9pt

\noindent
{\bf{Definition \ref{S:RealPlane}.1}}\ \ 
An analytic function $f_{\smallindex{W}}(z)$ is said to be the \emph{standard completion} of a 
\R2 harmonic function $W$, for some suitable common region in the $x$-$y$ plane, when $f_{\smallindex{W}}(z) \eq u_{\smallindex{W}}(x,\,y) + i\, v_{\smallindex{W}}(x,\,y)$, where $u_{\smallindex{W}}(x,\,y) \eq W(x,\,y)$ and $f_{\smallindex{W}}$ is the analytic completion of $u_{\smallindex{W}}(x,\,y)$ [i.e., $v_{\smallindex{W}}(x,\,y)$ is the unique complex function required to make $f_{\smallindex{W}}$ analytic].

\vskip 9pt

\noindent
After taking branch cuts into account, the standard completion is unique since the real and complex parts of an analytic function are separately harmonic and there is thus a one-to-one correspondence of harmonic and analytic functions \cite{Nehari} (for restrictions see Duren \cite{DurHar}).

   For dipole and higher order poles in \R2 this standard completion process seamlessly meshes with the  corresponding previous formalism given for pole strength determination in \C\ and it allows for easy implementations.  For example, a dipole aligned along the $x$-axis in \R2 is proportional to
\begin{equation}\label{E:LogW}
u_{\smallindex{W}} \eq -\frac{\partial \ }{\partial x} \ln \, \frac1{\sqrt{(x - x_k)^2 + (y -y_k)^2}} =
 \frac{x -x_k}{\sqrt{(x - x_k)^2 + (y -y_k)^2}}\ ,
\end{equation}
 which has a standard completion of
\begin{equation}
f_{\smallindex{W}} = \frac{x -x_k}{\sqrt{(x - x_k)^2 + (y -y_k)^2}} +  \frac{i\,(y -y_k)}{\sqrt{(x - x_k)^2 + (y -y_k)^2}} = \frac1{z - z_k}\ .
\end{equation}
  Likewise a dipole aligned along the $y$-axis in \R2 is proportional to
\begin{equation}
u_{\smallindex{W}} \eq -\frac{\partial \ }{\partial y} \ln \, \frac1{\sqrt{(x - x_k)^2 + (y -y_k)^2}}
\end{equation}
 which has a standard completion of $-i/(z - z_k)$.  Since linear superposition holds for this standard completion process, an arbritrary combination of various dipole terms can be easily handled by the resulting DIDACKS formalism.  Handling \R2 harmonic problems for the exterior of a disk thus presents no real conceptual difficulties since the usual procedure of mapping such problems into \C\ and then recovering the results by an inverse transformation can be easily applied, or one can map the functions and inner products obtained in the complex setting into the real \R2 setting by using relationships analogous to (\ref{E:Eip}) and (\ref{E:Rlog1}).
 
  The inclusion of point logarithmic terms in \R2 is not quite so trivial, but towards that end consider the \R2 potential term $L_k(x,\,y) \eq \ln\,[1/{\sqrt{(x - x_k)^2 + (y -y_k)^2}}]$.  The standard completion of $L_k(x,\,y) = \ln\,[1/{\sqrt{(x - x_k)^2 + (y -y_k)^2}}] = \ln\,(1/{|z - z_k|})$ \ is \ $\ell_k(z) = \ln\,[1/(z - z_k)] = -\ln\, (z - z_k)$, which was not considered in Section~\ref{S:SIDACKS} or Section~\ref{S:DIDACKS}.  Nevertheless, retracing the steps that lead up to (\ref{E:CollPropD}) for this logarithmic function is straightforward and yields the following suggestive relationship
\begin{equation}\label{E:LogGCP}
 (f,\,\ln [(z - z_k)^{-1}]){\ls}_{D}  \overset{F} = \frac12\sum\limits_{n=1}^{\infty} a_n^{*}z^n_k \eq \frac12\,f^{*}(p_k^{*})\ ;
\end{equation}
where the symbol $\overset{F} =$ indicates a formal manipulation (or relationship), which may or may not be fully justifiable.  It is easy to see that (\ref{E:LogGCP}) is not suitable as a framework for applications and then to discern the underlying reasons why.  Towards that end, first substitute $-\ln (z - z_{k'})$ for $f(z)$ in the left hand side of (\ref{E:LogGCP}) (where $k' \neq k$).  Next replace $k$ by $k'$ and then replace $f(z)$ by $-\ln (z - z_{k})$ in the left hand side of (\ref{E:LogGCP}).  Comparison of the net result for these two expressions yields $(-\ln (z - z_{k'}),\,-\ln (z - z_k))_{D} \neq (-\ln (z - z_{k}),\,-\ln (z - z_{k'}))^{*}_{D}$.  The difficulty here is that $\ln\,[1/{z - z_k}]$ is not in the class of admissible analytic functions as specified by (\ref{E:series1}).  Not only does $\lim_{|z| \to \infty} \ln (z - z_k) \neq 0$, but this limit is not even bounded.  Likewise, of course, $\ln\,[1/{\sqrt{(x - x_k)^2 + (y -y_k)^2}}]$ is not in the class of admissible functions for \R2.  The solution is simple.  Consider the complex case first.  Since the sum of two analytic functions is analytic, it is obvious that
\begin{equation}\label{E:xibasis}
 {\xi}_k(z) \eq \ln\,\frac1{z - z_k} - \ln\,\frac1{z - z'_k}\, = \,\ln\,\left(\frac {z - z'_k}{z - z_k}\right)
\end{equation}
satisfies the conditions (\ref{E:series1}).  It is thus suitable as a candidate basis function for logarithmic fits.  Here $z'_k$ is called the paired point of $z_k$ and it is an arbritrary complex number, except for the restrictions that $|z'_k| < 1$ and $z'_k \neq z_{k}$ for all $k$.  With these modifications (\ref{E:LogGCP}) can then be employed to obtain
\begin{equation}\label{E:xi}
(f,\,{\xi}_k){\ls}_{D} = \frac12\,f^{*}(p_k^{*}) - \frac12\,f^{*}((p'_k)^{*})\ ,
\end{equation}
where $(p'_k)^{*}\,$ is the conjugate involution of the paired point of $z_k$ and is thus specified by taking the conjugate of $p'_k \eq 1/{z'_k}$.
Equation~(\ref{E:xi}) is self-consistent and it can be used to directly verify that
$({\xi}_k,\,{\xi}_{k'})^{*}_{D} = ({\xi}_{k'},\,{\xi}_k)_{D}$ as expected.  It is clear that ${\xi}_k$ can be used to perform various types of fits in the same way that the basis functions $1/(z - z_k)$ were used; however, the choice of paired points clearly affects the resulting quality of the fit.  For example, when the values of $z_k$ are determined by a NLLSQ process it would be advantageous to also determine the paired point locations as part of the same NLLSQ process, although this clearly complicates the implementation.  Even for a LLSQ fit, the introduction of paired points is often inconvenient since it effectively doubles the number of point logarithmic sources and the amount of point evaluation information that is required for $f$.   Thus generally it is convenient to consider another set of candidate basis functions given by
\begin{equation}\label{E:logbasis}
 {\psi}_k(z) \eq  \,\ln\,\frac1{z - z_k} - \,\ln\,\frac1{z}\, = \,\ln\,\frac z{(z - z_k)} \ ,
\end{equation}
where all the paired points have effectively been combined and situated at the origin.  When $z'_k \eq 0$, the paired conjugate involution points are at infinity and the second term on the right hand side of (\ref{E:xi}) vanishes for all $k$, by convention, and thus
\begin{equation}\label{E:psiDACK}
(f,\,{\psi}_k){\ls}_{D} = \frac12\,f^{*}(p_k^{*})\ .
\end{equation}
Clearly one can handle logarithmic source fits to analytic functions just as before by using the basis functions ${\xi}_k$ or ${\psi}_k$; however, the branch cuts associated with these logarithmic forms raise certain issues---especially when $f(z)$ does not have them.  Notice that these basis functions are not translation invariant whereas the basis functions ${\xi}_k(z)$ specified by (\ref{E:xibasis}) are.

   When logarithmic fits to a harmonic function $W(x,\,y)$ in \R2 are considered these branch-cut concerns clearly no longer apply, but the requirement that the basis functions tail off quickly enough to zero at infinity still remains.  Keeping this concern in view, consider the following set of basis functions 
\begin{equation}\label{E:PsiK}
{\Psi}_k = \ln\,\frac1{|{\vec{X}} - {\vec{X}}_k|} - \ln\,\frac1{|{\vec{X}}|} =  \ln\,\frac{|{\vec{X}}|}{|{\vec{X}} - {\vec{X}}_k|}\ ,
\end{equation}
where ${\vec{X}} \eq (x,\,y)^T$ and ${\vec{X}}_k \eq (x_k,\,y_k)^T$ satisfy $|\vec{X}| \geq 1$ and $0 < |{\vec{X}}_k| < 1$.  Let $U = U(\vec{X})$ denote the linear superposition of these logarithmic basis terms which will be used to approximate $W$, then 
\begin{equation}\label{E:Ulog}
U(\vec{X}) = \sum\limits_{k=1}^{N_k} {m}_k\,{\Psi}_k({\vec{X}}) ,
\end{equation}
where $m_k \in \mRR$.  

  On comparing (\ref{E:PsiK}) and (\ref{E:logbasis}) it is obvious that the standard completion of ${\Psi}_k({\vec{X}})$ is ${\psi}_k(z)$ and thus that the standard completion of $U$ is the analytic function $f_{\smallindex{U}}(z)$ where
\begin{equation}\label{E:Ucomp}
  f_{\smallindex{U}}(z) \eq \sum\limits_{k=1}^{N_k} {\mu}_k\, {\psi}_k(z) 
\end{equation}
and, in general, ${\mu}_k \in \mC$.  In common practice \R2 harmonic problems are often mapped to \C\ and back using standard harmonic completion to take advantage of various techniques available in the complex domain (such as conformal mapping).  In the present context, when this ``round-trip mapping'' is considered for actual applications several interpretational issues immediately arise in addition to those associated with branch cuts. First there is the minor one involving the meaning of complex values of ${\mu}_k$.  Leaving this issue aside, another more fundamental one arises from the act of standard completion of $W$ itself, since there is insufficient given information to insure that this completion is unique.  In particular, a DIDACKS fit utilizes only information about $W$ at specified points in \R2 and this by itself is obviously insufficient to uniquely specify the form of $W$ over the whole region of interest so that various harmonic completions are theoretically possible.  Specifically, at the specified data points diverse analytic functions can have identical real parts that match $W$.  While a minimum principle clearly underlies this standard completion process for DACKS fits, it is unclear whether there may be some sort of unexpected  interaction with this minimum principle involving the possibility of complex values of ${\mu}_k$ and branch cuts.  It is thus at least conceptually desirable to quantitatively analyze this problem.  This can be best accomplished by proving that the results of a fit performed in the complex setting must match the same fit performed in the real \R2 setting, but to do this requires that a unique corresponding real setting theory be available.  Moreover, if this real setting theory exists it far easier in both a practical and theoretical sense to simply use it and side-step this round-trip mapping altogether.  Fortunately, developing a corresponding theory based solely on the real Dirichlet form $\D[f,\,g]$ is fairly easy.  The basic strategy will thus be to map the underlying inner products themselves from \C\ to \R2 (which only needs to be done once) rather than map the functions from \R2 to \C\ and back for every application.  (This is clearly the most desirable way to proceed in many other situations encountered in complex variable/\R{2} potential theory as well.)

  Let $G(\vec{X})$ and $H(\vec{X})$ be two functions as specified from the context of (\ref{E:ReDirInt}) and let $h(z)$ and $g(z)$ be their corresponding standard completions.  Substituting
\begin{equation}
 \frac{d\,g}{d z} = \frac{\partial\,G}{\partial\,x} + i\,\frac{\partial\,G}{\partial\,y}\ \ {\text{and}}\ \   \frac{d\,h}{d z} = \frac{\partial\,H}{\partial\,x} + i\,\frac{\partial\,H}{\partial\,y}\ 
\end{equation}
into (\ref{E:CplxDirInt}) shows that
\begin{equation}\label{E:RCdir}
\D[G,\,H] = \text{Re}\,\{\D_c[g,\,h]\}
\end{equation}
and thus
\begin{equation}\label{E:Eip}
(G,\,H){\ls}_{E} = {\text{Re}}\,\{\,(g,\,h){\ls}_{D}\}\ , 
\end{equation}
where
\begin{equation}\label{E:energy}
 (G,\,H){\ls}_{E} \eq \frac1{2\pi}\,\D[G,\,H] 
\end{equation}
defines the energy inner product.  Here, aside from a units dependent constant, the energy associated with the field ${\vec{E}} = -\nabla W$ is given by $\|\,W\,\|^2_E \eq {(W,\,W)_E\,}$.  In order to perform various types of \R2 DIDACKS fits, all that is left to do is to determine closed form expressions for $(G,\,K_D)_E$, where $K_D = K_D(\vec{X},\,{\vec{X}}_k)$ is any suitable DIDACK.  Using (\ref{E:Eip}), it is clear that most of the work has already been done in the form of (\ref{E:CollPropD}), (\ref{E:LogGCP}) and (\ref{E:psiDACK}).  For example, since ${\psi}_k(z)$ is the standard completion of ${\Psi}_k({\vec{X}})$, (\ref{E:psiDACK}) immediately yields
\begin{equation}\label{E:Rlog1}
(G,\, {\Psi}_k){\ls}_{E} = \frac12\,{\text{Re}}\,\{(g(p_k^{*}))^{*}\} \ .
\end{equation}
Further, recalling that $G(\vec{X}) = {\text{Re}}\,\{g(z)\}$ allows for the reinterpretation of the right hand side of (\ref{E:Rlog1}), but the occurrence of $p_k^{*}$ here is perhaps disconcerting at first. Notice however that ${\text{arg}}\,\{z_k\} = {\text{arg}}\,\{p^{*}_k\} = {\theta}_k$, so that in the polar representation ${\text{Re}}\,\{(g(p_k^{*})^{*}\} = G(1/r_k,\,{\theta}_k)$.  Introducing ${\vec{P}}_k \eq {\vec{X}}_k/|{\vec{X}}_k|^2$ allows (\ref{E:Rlog1}) to be reexpressed as 
\begin{equation}\label{E:Rlog}
(G,\, \ln\, (|\vec{X}|/|\vec{X} - {\vec{X}}_k|)){\ls}_{E} = \frac12\, G({\vec{P}}_k)\ .
\end{equation}
Similar higher order pole expressions can be obtained directly from (\ref{E:Rlog}) by taking partials on both sides with respect to $x_k$ or $y_k$ and recalling, for example, that
\begin{equation}\label{E:grads}
\frac{\partial\ }{\partial\,x_k}\,\frac1{|\vec{X} - {\vec{X}}_k|} = - \,\frac{\partial\ }{\partial\,x}\,\frac1{|\vec{X} - {\vec{X}}_k|}\ . 
\end{equation}

   Basing fits on the expansion (\ref{E:Ulog}) and it generalizations is thus straightforward since partials of ${\Phi}_E \eq \| W - U||_E^2$ with respect to $m_k$ (or other source terms) are be easily determined and, as before, when the results are set to zero closed-form linear equation sets result that can be easily solved.  Obviously (\ref{E:grads}) implies that fits based on specific values of $\mathbf{\nabla} W$ at selected points are dipole fits; moreover, similar correspondences between higher order poles (quadrupoles, etc.) and their associated point measurement values hold.

  Various other possibilities clearly exist besides those outlined in the preceeding sections.  For example, in the complex setting, based on the fact that $1/(z - z_k)$ is a shared replicating kernel for the standard integral norm and $D$ norm, an alternative DACKS norm can be formed by taking a linear combination of these two norms: $(f,\,g)_D + \lambda (f,\,g)_{\sigma}$\,, where $\lambda$ is real and positive.  Another possibility arises from the fact that a single norm may have two (or more) different replicating kernels: thus in the complex setting an alternative DIDACK arises from the linear superposition of $(z - z_k)^{-1}$ and $\ln\,[ z/{(z - z_k)}]$ [c.f., (\ref{E:CollPropD}), (\ref{E:logbasis}) and (\ref{E:psiDACK})].  Although this and other such possibilities are interesting in themselves, they will not been addressed in the sequel.

\section{Numerical Tests}\label{S:examples}
  
   This section presents numerical test results from various SIDACKS and DIDACKS implementations.  The emphasis here is on the complex setting and, in fact, real \R2 results will not be considered until the occurrence of (\ref{E:nbasis}) towards the end of the section. 

    For each implementation there are three primary aspects that must be considered:
\begin{enumerate}
\item
  The choice of basis functions. 
\item
The choice of norm. 
\item
The nature of the function to be approximated.
\end{enumerate}
The third item on this list will be considered in more detail after computational issues associated with the first two items are fully addressed.  Thus while the choices of basis functions and norms considered here, as well as their merits and the various issues involved, have been previously discussed and thus it is generally only necessary to take note of particular usages in the sequel and then observe the quality of the resulting fits; there are, however, associated specific computational issues that merit further discussion.  In particular, the numerical nature of the linear equation set to be solved is clearly depend on the choice of norm and basis functions.  Since, in all cases, (\ref{E:TmuA}) characterizes the resulting linear equation sets, it is useful to have some suitable measure of the computational stability of this equation set or of the inverse of the matrix $\mathbf{T}$ that occurs there.  The condition number of $\mathbf{T}$ is used for this purpose and it is taken to be the ratio of the largest to the smallest eigenvalue of $\mathbf{T}$.  (Since all of the eigenvalues encountered here are positive real numbers, the absolute value need not be taken here.)

   It is worth noting in the sequel that while the results are good, condition numbers for particular test cases are often surprisingly large.  As discussed below all three of the factors mentioned above have a bearing on this phenomenon, but the effects due to basis functions clearly dominate since with a choice of orthonormal basis functions the computationally trivial situation $\mathbf{T} = \mathbf{I}$ results (where  $\mathbf{I}$ is the identity matrix and the condition number is thus $1$).  This is one of three interconnected aspects to be considered in conjunction with the basis functions dealt with here:  their spatial localizability, their frequency spread and their non-orthogonality.  By way of concrete example, consider the basis functions $1/(z - z_k)$ in conjuction with the SIDACKS treatment of  Section~\ref{S:SIDACKS}.  As $|z_k|$ approaches $1$ this basis function is clearly suitable for matching very localized effects that occur on the unit circle, but in more common situations non-localized effects must be matched.  In this sense the properties of the function to be approximated matter a great deal:  if only a few non-overlapping localized effects on the unit circle are to be matched then a very well conditioned set of sources with $|z_k| \approx 1$ can be utilized.  It is primarily when there is a sizeable low-frequency content in the function to be approximated that large condition numbers are naturally encountered.  From a frequency perspective, when specialized to the case $r = 1$, (\ref{E:geoseries}) yields a Fourier-like series that has coefficients with a magnitude of $|z_k|^{n-1}$ [c.f., (\ref{E:RealFser})]---which is clearly very spread out in the frequency domain unless $|z_k|$ is small.  While it is a property of low frequency functions to be spread out in the spatial domain, basis functions of the form $1/(z - z_k)$ have a great deal of spatial overlap for $|z_k| \approx 0$ and this directly implies a high degree of non-orthogonality or intercorrelatedness that, in turn, is responsible for the large condition numbers encountered here.  As discussed at the end of Section~\ref{S:PreDIDACKS}, the norms themselves also play a role here since they fail to exert a direct control over the source coefficients themselves.  The reasons for this are linked to underlying aspects of the modeling philosophy itself and are worth emphasizing.  For modeling problems the criteria for success is fairly direct and is built explicitly into the definition of the norms themselves and it is these norms that are minimized.  Thus, while the error in the coefficients associated with a particular fit may not be well controlled, this generally does not matter since these errors do not show up in the norms.  Moreover, for a pure modeling problem it is easy to validate a candidate fit at any number of points and directly ascertain its suitability.  Hence for modeling problems condition number considerations only matter when they directly effect the quality of the fit itself and this often occurs only as a second-order effect when machine round-off effects become significant---provided, of course, that the sources are located at appropriate points. 

   Several specific computational details with regards to the solution of (\ref{E:TmuA}) and condition numbers are worth addressing.  Obviously, to obtain reasonable results in the presence of large condition numbers some care with regards to computational word length and equation solving techniques are required.  Toward this end all standard complex variable computations were performed using a 64-bit word length, which corresponds to about 16 significant digits on the system used (on many systems 64-bit words have one or two less significant digits).  These complex number evaluations were then mapped to and from real linear equation sets via (\ref{E:realT}), which results in each eigenvalue being duplicated (as discussed below).  These real linear equations themselves were solved using Householder triangulation implemented with a 128-bit word length (33-34 significant digits).  (If the system architecture makes use of the fact that part of the original word length is reserved for the mantissa---as the system employed does---then it is possible to obtain more than double the number of significant digits by doubling the bit size.) The eigenvalues and thus the condition numbers themselves were computed using singular value decomposition (SVD) software implemented with the same 128-bit word length.  (All the computed eigenvalues encountered were positive as expected.)

  When testing the modeling application algorithms proposed here, it is useful to consider the behavior that might be encountered with regards to not only various basis functions and norms, but also with regards to various types of functions that might be fit. First, the specified class of admissible functions---which is somewhat restricted---must always be kept in mind.  This is not a particular issue with respect to basis functions since all of the basis functions considered here satisfy the needed requirements, but since $f$ is usually an externally  specified function, it is not necessarily a given that it can be recast in a suitable form, even after extensive modifications.   Consider, for example, the form $f(z) =1/\sqrt{z} = \sqrt{r^{-1}}e^{-i\theta/2}$.  Here $(f(z) - a_0)$ cannot be expressed in the form of (\ref{E:series1}) even though it satisfies the Cauchy-Riemann conditions in polar form for all $1 < |z| < \infty$, since it is obvious that $1/\sqrt{z}$ is not analytic at $|z| = \infty$, just as $\sqrt{z}$ is not analytic at $z = 0$.  (To the contrary, suppose that a series of the form $1/\sqrt{z} = \sum_{n=0}^{\infty} a_n\,z^{-n}$ exists, then squaring both sides of this form and identifying like powers of $1/z$ yields an inconsistent result.)  Clearly the requirement that candidate functions to be fit satisfy the demands of (\ref{E:series1}) is somewhat restrictive and, in fact, many standard  analytic functions defined in the exterior of a unit disk do not satisfy this criteria.  For example, $\sin(z)$ is obviously analytic and bounded as $|z| \rightarrow \infty$; however, it does not satisfy the necessary criteria since $\lim_{\,\,\ \ |z| \to \infty}\, f(z) \rightarrow 0$ fails to hold; moreover, even subtraction of a constant fails to produce an acceptable function.  Clearly when a LLSQ SIDACKS or DIDACKS fit is performed to a function $f$ that is not in the class of admissible functions then large errors will (generally) result since, although the replication properties at the data points are enforced, the minimum norm process will be satisfied not with regards to the specified $f$ as intended, but rather with respect to some function that satisfies the same replication constraints and that is in the class of admissible functions.  In such cases some other approximating procedure should clearly be considered.  Nevertheless, it is not difficult to come up with suitable functional forms for testing that are based on standard analytic functions, which may not themselves be in the class of acceptable functions.  The last three of the six test functions to be fit were chosen with this consideration in mind.  Since the discussion of the three items above is complete, the full list of test functions and other implementation detail will be considered next. 

 Fits of various types were performed to one of six specified test functions, denoted $f_j$ for $j = 1,\,2,\,3,\,4,\,5,\,6$ and given by:
\begin{itemize}
\item
$f_1 \eq C_1\sum\limits_{n=1}^{5} z^{-n}\ \ \text{where}\ \ \, C_1 \eq \frac{{1 + i}}{2\sqrt{n}}$
\item
$f_2 \eq C_2\,\sum\limits_{n=1}^4 \,\frac1{(z - z_n)}\ \ \text{where}\ \ \, C_2 \eq {(1 +i)}/{6}\ \ \text{and}\ \, z_n \eq (3/4)\,e^{i\,\pi(n-1)/2}$ 
\item
$f_3 \eq C_3\sum\limits_{n=1}^4 \,\ln\,\left[ z/{(z - z_n)}\right]\ \ \text{where}\ \ \, C_3 \eq 2(1 + i)\ \ \text{and}\ \,z_n \eq (3/4)\,e^{i\,\pi(n-1)/2}$ 
\item
$f_4 \eq \sin (1/z)$
\item
$f_5 \eq 2[\cos(1/z) - 1]$
\item
$f_6 \eq e^{1/z} - 1$ 
\end{itemize}
The scale of each of these test functions was chosen so that the resulting approximation errors  would be more-or-less comparable.  Table~\ref{Ta:SpecifiedFs} summarizes data that gives a feeling for the properties of these various test functions.  It was obtained by computing the magnitude of each $f_j$ at 1,000 uniformly spaced points on the unit circle.
\begin{table} 
\begin{center}
\begin{tabular}{|c|c|c|c|c|c|c|}\hline
  Function Type    & $f_1$ & $f_2$ & $f_3$ & $f_4$ & $f_5$ & $f_6$  \\\hline
 Average Magnitude  & 0.87 & 0.97 & 0.90 & 1.01 & 1.00 & 1.01\\\hline
 Maximum Magnitude & 2.23  & 1.38 & 1.08 & 1.18 & 1.09 & 1.72\\\hline
 $\|\,f_j\,\|_{\sigma}$ &1.07&0.99&0.91&1.01&1.00&1.13\\\hline
\end{tabular}
\caption{Summary Data for Specified Functions on Unit Circle}\label{Ta:SpecifiedFs}
\end{center}
\end{table}

\vskip .02in
\begin{table} 
\begin{center}
\begin{tabular}{|c|c|c|c|c|c|c|}\hline
Case  &      & Function  & \multicolumn{4}{c|}{Basis Function Specification} \\\cline{4-7} 
\raisebox{1.4ex}[0cm][0cm]{Number} & \raisebox{2.7ex}[0cm][0cm]{Norm} & \raisebox{1.4ex}[0cm][0cm]{to be Fit} & Function & $R_B$ & $N_k$ & $1/z$ Term? \\ \hline\hline
 1      &  $D$     & $f_1$ & $1/(z\!-\!z_k)$ & 0.6  &  8 & No  \\ \hline
 2      &  $D$     & $f_1$ & $1/(z\!-\!z_k)$ & 0.6  & 16 & No  \\ \hline
 3      &  $D$     & $f_1$ & $1/(z\!-\!z_k)$ & 1/2  & 16 & No  \\ \hline
 4      &  $D$     & $f_1$ & $1/(z\!-\!z_k)$ & 0.27 & 16 & No  \\ \hline
 5      & $\sigma$ & $f_1$ & $1/(z\!-\!z_k)$ & 1/2  & 16 & No  \\ \hline
 6      &  $D$     & $f_1$ & ${\psi}_k $     & 1/2  & 16 & No  \\ \hline
 7      &  $D$     & $f_2$ & $1/(z\!-\!z_k)$ & 1/2  & 16 & No  \\ \hline
 8      &  $D$     & $f_2$ & $1/(z\!-\!z_k)$ & 0.7  & 16 & No  \\ \hline
 9      &  $D$     & $f_3$ & $1/(z\!-\!z_k)$ & 1/2  & 16 & No  \\ \hline
 10     &  $D$     & $f_3$ & $1/(z\!-\!z_k)$ & 0.7  & 16 & No  \\ \hline
 11     &  $D$     & $f_2$ & ${\psi}_k $     & 1/2  & 16 & No  \\ \hline
 12     &  $D$     & $f_2$ & ${\psi}_k $     & 0.7  & 16 & No  \\ \hline
 13     &  $D$     & $f_3$ & ${\psi}_k $     & 1/2  & 16 & No  \\ \hline
 14     &  $D$     & $f_3$ & ${\psi}_k $     & 0.7  & 16 & No  \\ \hline
 15     &  $D$     & $f_4$ & $1/(z\!-\!z_k)$ & 1/2  & 16 & No  \\ \hline
 16     &  $D$     & $f_4$ & ${\psi}_k $     & 1/2  & 16 & No  \\ \hline
 17     &  $D$     & $f_5$ & $1/(z\!-\!z_k)$ & 1/2  & 16 & No  \\ \hline
 18     &  $D$     & $f_5$ & ${\psi}_k $     & 1/2  & 16 & No  \\ \hline
 19     &  $D$     & $f_6$ & $1/(z\!-\!z_k)$ & 1/2  & 16 & No  \\ \hline
 20     &  $D$     & $f_6$ & ${\psi}_k $     & 1/2  & 16 & No  \\ \hline
 21     & $\sigma$ & $f_1$ & $1/(z\!-\!z_k)$ & 1/2  & 17 & Yes \\ \hline
 22     &  $D$     & $f_1$ & $1/(z\!-\!z_k)$ & 1/2  & 17 & Yes \\ \hline
 23     &  $D$     & $f_1$ & ${\psi}_k $     & 1/2  & 17 & Yes \\ \hline
\end{tabular}
\caption{Case Number Specifications}\label{Ta:CaseData}
\end{center}
\end{table}

   For testing purposes single ring configurations of a specified source type were used as basis function sets and, as mentioned earlier, only the complex setting was considered first. (These choices were made primarily for convenience.)  The basis function radius, $R_B$, is shown for various specific test case configurations in Table~\ref{Ta:CaseData}.  The basis functions used were either $1/(z\!-\!z_k)$ or ${\psi}_k $ [c.f., (\ref{E:logbasis})], as also indicated in Table~\ref{Ta:CaseData}.  For both sets of basis functions their locations $z_k$ were specified by 
\begin{equation}\label{E:zk}
z_k \eq R_B\,e^{i\,2\pi(k-1)/{N_k}}\ \ \text{where}\,\ R_B \in {\mRR}\ \ \text{and}\,\ 0 < R_B < 1.
\end{equation}
Here $N_k$, the number of basis functions used, is also indicated in Table~\ref{Ta:CaseData}, as is whether the complex energy DIDACKS norm ($D$) or the SIDACKS norm ($\sigma$) was used.  Finally for each case the function to be approximated is also indicated in Table~\ref{Ta:CaseData} so that it contains all of the relevant data for the test configurations and associated case numbers, but it does not contain the results.  The results for each of the cases specified in Table~\ref{Ta:CaseData} is presented in Table~\ref{Ta:CaseResults}.  The results of each fit were evaluated at 1,000 uniformly spaced points on the unit circle ($R_E = 1$) and on a circle of twice that radius ($R_E = 2$).

  A comparison of the case descriptions for cases~1 \& 2 given in Table~\ref{Ta:CaseData} with the results specified by Table~\ref{Ta:CaseResults} indicates that when the number of fitting points ($N_k$) is increased the accuracy improves, as expected. (If the added points are at irregular and inappropriate locations, this well may not be the case---especially for large  condition numbers.)  Comparison of the results for cases~2 through 4 indicate that as $R_B$ becomes smaller, the error decreases---but there are limits. (Below $R_B = 0.27$ the standard deviation may be smaller, but not consistently so---indicating the onset of unwanted machine round-off effects.)  Comparison of cases~3 \& 5 indicates that the $\sigma$ and $D$ norms give roughly the same accuracy at $R_E = 1$. Comparison of cases~3, 7, 9, 15, 17 and 19 shows that the approach works well for various types of more or less smooth functions. A comparison of cases~6, 11, 13, 16, 18 and 20 shows that the same conclusion holds for logarithmic basis functions as well as for simple poles.  As expected from the above discussion, comparison of cases~7 \& 8 and/or 9 \& 10 shows that smaller values of $R_B$ may not be appropriate for certain functions and, in fact, that the choice of source depth ($R_B$) and spacing (i.e., $N_k$) may be highly dependent on the nature of the function to be fit.  Cases~11 through 14 show that this conclusion is not particularly dependent on the choice of basis function type.  In this connection, notice that in some sense $f_2$ and $f_3$ are more irregular than $f_1$ since the series representations of $f_2$ and $f_3$ in the form (\ref{E:series1}) do not terminate.  Functions $f_4$, $f_5$ and $f_6$ are also very well behaved and the effects of basis function choice for each of these function types can be had through a study of cases 14 through 20.  Finally the inclusion of a single additional pole source at the origin (i.e., $1/z$) improves accuracy, but raises the condition number, as is apparent from comparison of cases 21 through 23 with cases~5, 3 and 6 respectively.  Computationally, for the $\sigma$ norm these $1/z$ terms are easy to include since it is only necessary to add an appropriate row and column to the matrix (\ref{E:TmuA}).  For example, a direct substitution of $1/z$ and the series representation (\ref{E:series1}) into (\ref{E:sigg1}) yields the needed additional rows and columns to fill out the $\mathbf{T}$ matrix:  $(1/z,\,(z - z_k)^{-1})_{\sigma} = ((z - z_k)^{-1},\,1/z)_{\sigma} =
(1/z,\,1/z)_{\sigma} = 1$.  Also when a row is added to $\mathbf{A}$ the additional element is $A_{N_k+1} = a_1$.  The various required terms for the $D$ norm are obtained in a like fashion.

\vskip .02in
\begin{table} 
\begin{center}
\begin{tabular}{|c|c|c|c|c|c|}\hline
 &     & \multicolumn{2}{c|}{Fit Errors at $R_E = 1$} & \multicolumn{2}{c|}{Fit Errors at $R_E = 2$}\\\cline{3-6} 
 \raisebox{1.6ex}[0cm][0cm]{Case} & \raisebox{1.6ex}[0cm][0cm]{Condition}  & Standard & Maximum & Standard & Maximum \\ 
 \raisebox{2.6ex}[0cm][0cm]{Number}& \raisebox{2.6ex}[0cm][0cm]{Number} & \raisebox{1.4ex}[0cm][0cm]{Deviation} & \raisebox{1.4ex}[0cm][0cm]{Magnitude} & \raisebox{1.4ex}[0cm][0cm]{Deviation} & \raisebox{1.4ex}[0cm][0cm]{Magnitude} \\ \hline\hline
 1      &  .160$\times 10^3$  &.180$\times 10^{-1}$  & .377$\times 10^{-1}$ & .916$\times 10^{-3}$ &.116$\times 10^{-2}$   \\ \hline
 2      &  .283$\times 10^6$  &.302$\times 10^{-3}$  & .643$\times 10^{-3}$ & .488$\times 10^{-6}$ &.605$\times 10^{-6}$   \\ \hline
 3      &  .671$\times 10^8$  &.163$\times 10^{-4}$  & .348$\times 10^{-4}$ & .143$\times 10^{-8}$ &.189$\times 10^{-8}$   \\ \hline
 4      &.701$\times 10^{16}$ &.123$\times 10^{-8}$  & .223$\times 10^{-8}$ &.105$\times 10^{-12}$ &.118$\times 10^{-12}$ \\ \hline
 5      &.107$\times 10^{10}$ &.163$\times 10^{-4}$  & .349$\times 10^{-4}$ & .125$\times 10^{-9}$ &.257$\times 10^{-9}$   \\ \hline
 6      &.172$\times 10^{11}$ &.214$\times 10^{-5}$  & .469$\times 10^{-5}$ & .881$\times 10^{-11}$ &.225$\times 10^{-10}$ \\ \hline
 7      &.671$\times 10^{8} $ &.995$\times 10^{-2}$  & .138$\times 10^{-1}$ & .123$\times 10^{-5}$ &.130$\times 10^{-5}$   \\ \hline
 8      &.277$\times 10^{4} $ &.667$\times 10^{-2}$  & .889$\times 10^{-2}$ & .179$\times 10^{-3}$ &.179$\times 10^{-3}$   \\ \hline
 9      &.671$\times 10^{7} $ &.185$\times 10^{-2}$  & .243$\times 10^{-2}$ & .866$\times 10^{-8}$ &.103$\times 10^{-7}$   \\ \hline
 10     &.277$\times 10^{4} $ &.118$\times 10^{-2}$  & .122$\times 10^{-2}$ & .123$\times 10^{-5}$ &.123$\times 10^{-5}$   \\ \hline
 11     &.172$\times 10^{11}$ &.997$\times 10^{-2}$  & .138$\times 10^{-1}$ & .102$\times 10^{-6}$ &.146$\times 10^{-6}$   \\ \hline
 12     &.710$\times 10^{6} $ &.971$\times 10^{-2}$  & .132$\times 10^{-1}$ & .154$\times 10^{-4}$ &.157$\times 10^{-4}$   \\ \hline
 13     &.172$\times 10^{11}$ &.186$\times 10^{-2}$  & .244$\times 10^{-2}$ & .242$\times 10^{-8}$ &.349$\times 10^{-8}$   \\ \hline
 14     &.710$\times 10^{6} $ &.124$\times 10^{-2}$  & .163$\times 10^{-2}$ & .249$\times 10^{-6}$ &.253$\times 10^{-6}$   \\ \hline
 15     &.671$\times 10^{7} $ &.155$\times 10^{-4}$  & .179$\times 10^{-4}$ & .198$\times 10^{-8}$ &.213$\times 10^{-8}$   \\ \hline
 16     &.172$\times 10^{11}$ &.984$\times 10^{-6}$  & .133$\times 10^{-5}$ & .975$\times 10^{-11}$ &.151$\times 10^{-10}$ \\ \hline
 17     &.671$\times 10^{8} $ &.153$\times 10^{-4}$  & .166$\times 10^{-4}$ & .527$\times 10^{-9}$ &.589$\times 10^{-9}$   \\ \hline
 18     &.172$\times 10^{11}$ &.172$\times 10^{-5}$  & .196$\times 10^{-5}$ & .916$\times 10^{-11}$ &.134$\times 10^{-10}$ \\ \hline
 19     &.671$\times 10^{8} $ &.173$\times 10^{-4}$  & .262$\times 10^{-4}$ & .200$\times 10^{-8}$ &.242$\times 10^{-8}$   \\ \hline
 20     &.172$\times 10^{11}$ &.131$\times 10^{-5}$  & .231$\times 10^{-5}$ & .108$\times 10^{-10}$ &.218$\times 10^{-10}$ \\ \hline
 21     &.776$\times 10^{11}$ &.122$\times 10^{-4}$  & .241$\times 10^{-4}$ & .452$\times 10^{-10}$ &.955$\times 10^{-10}$ \\ \hline
 22     &.456$\times 10^{10}$ &.122$\times 10^{-4}$  & .241$\times 10^{-4}$ & .276$\times 10^{-9}$ & .417$\times 10^{-9}$  \\ \hline
 23     &.456$\times 10^{10}$ &.204$\times 10^{-5}$  & .406$\times 10^{-5}$ & .554$\times 10^{-11}$ & .131$\times 10^{-10}$\\ \hline
\end{tabular}
\caption{Fit Results for Cases Specified in Table \ref{Ta:CaseData}}\label{Ta:CaseResults}
\end{center}
\end{table}

\vskip .01in

   Obviously the condition numbers given in Table~\ref{Ta:CaseResults} are somewhat large.
 Since the matrix $\mathbf{T}$ occuring in (\ref{E:sigLLSQ}) or (\ref{E:DLLSQ}) is Hermitian, the associated eigenvalues are real.  Further, since the norm of any function here is positive semidefinite and $\mathbf{T}^{-1}$ exists by assumption all of the $\mathbf{T}$ matrices encountered in testing were positive, it follows that $|\mathbf{T}| = \prod_k^{N_k}{\lambda}_k > 0$ holds, where the ${\lambda}_k$ are the $N_k$ eigenvalues of $\mathbf{T}$.  As previously mentioned, each of the eigenvalues ${\lambda}_k$ is repeated twice for the real symmetric matrix occuring on the left hand side of (\ref{E:realT}) and thus, when $\mathbf{T}$ is represented in this form, $|\mathbf{T}|$ can be computed by taking the square root of the product of all of its eigenvalues.  Since a SVD analysis of the symmetric real valued form of $\mathbf{T}$ was performed in all cases, the results of this determinant computation were readily available.  For SIDACKS cases this square root computation can be directly compared to the complex number evaluation of the right hand side of (\ref{E:TDet}).  This comparison was performed for all the relevant cases in Tables~\ref{Ta:CaseData} and \ref{Ta:CaseResults} and the results matched to within the expected accuracy.  For example, Case~5 yielded .1044048746849805$\times 10^{-52}$ for the square root eigenvalue computation versus .1044048714876914949$\times 10^{-52} - .52161\times 10^{-68}\,i$ from (\ref{E:TDet}).

  Finally, limited testing based on (\ref{E:PsiK}) and (\ref{E:Rlog}) was performed in the real \R2 setting.
 Similar results to those found in Table~\ref{Ta:CaseResults} were obtained.  Results from a single case here will have to suffice.  The function to be approximated was defined as
\begin{equation}\label{E:nbasis}
 F(\vec{X}) \eq \sum\limits_{n=1}^4 8(-1)^n\,\widetilde{\Psi}_n\,\ {\text{with}}\
\widetilde{\Psi}_n \eq \ln\, \frac{|\vec{X}|}{|\vec{X} - {\vec{X}}'_n|}\,\,,
\end{equation}
where ${\vec{X}}'_n \eq \frac13 (\,\cos {\delta}'_n,\,\sin {\delta}'_n)^T$ and where 
${\delta}'_n \eq (n-1)\pi/2$.  When evaluated at $r = 1$ this specified function definition yielded Max$\,\{\ |F(\vec{X})|\} = 1.785$ and Avg$\,\{|F(\vec{X})|\} = 1.130$. A real linear combination of 16 basis functions specified by (\ref{E:PsiK}) were used as an approximating form, where ${\vec{X}}_k \eq (2/5)\, (\,\cos {\Delta}_k,\,\sin {\Delta}_k)^T$ and ${\Delta}_k \eq (n-1)\pi/8$.  The condition number for the resulting $\mathbf{T}$ matrix was $.6939\times 10^{13}$.  Evaluated at $R_E = 1$, the resulting magnitude of the fit errors had a standard deviation of $.4124\times 10^{-4}$ and a maximum  of $.6053\times 10^{-4}$.  The corresponding errors at $R_E = 2$ were a standard deviation of $.4342\times 10{-7}$ and a maximum of $ = .759784\times 10^{-7}$.  Observe that since the fitting parameters are real there are half as many parameters as there are for the typical fits given in Table~\ref{Ta:CaseResults}.  Although the condition number was large here, the smallest eigenvalue was an outlier and excluding it in the inversion process (which can be easily done with Householder or SVD software) produced a fit error with almost identical errors that had a condition number of $.1490\times 10^{7}$.  While this is obviously one way to circumvent high condition numbers, it is also interesting that due to the smoothness of $F$ and the large word length (128-bit) good fits here can be obtained even for small values of $R_B$ (even below $.01$).  (Using $R_B = .01$ in the above basis function specifications and fitting to the same $F$ yields a fit with a standard deviation of $.3834\times 10^{-4}$ and a maximum of $ = .5469\times 10^{-4}$ at $R_E = 1$.  The condition number was $.7517\times 10^{31}$.)

\section{DACKS Theory for Disk Interiors}\label{S:IntDACKS}

  Finally SIDACKS and DIDACKS theory so far has focused exclusively on the exterior of a unit disk, but these DACKS approaches can easily be recast in terms of unit disk interiors. 
For convenience, it is useful to reuse the same symbols as much as possible, so it will be assumed in the sequel that $\infty > |z_k| > 1 \geq |z|$ specifies the basic parameters of interest for the interior case.  With this understanding, the inner product specified by (\ref{E:sigg1}) can be utilized as is with an appended subscript ${\sigma}'$ to replace the $\sigma$ subscript---where, as usual, the prime will be used to connote a complimentary region.  (While the compliment of the unit circle is the unit circle, the analytic region of interest is different for $\sigma$ and ${\sigma}'$.)  Further, it is useful to specify the admissible analytic functions for this interior geometry by upper case letters rather than lower case letters in order to keep in mind that they have a different series representation.  Thus consider a given function $F(z)$ that is to be approximated by a linear combination of simple poles located in some proper subset of the disk's exterior and which have the form $1/(z_k - z)$.  (The reverse sign convention is taken here so that $1/(z_k - z)$ will be positive when $z_k$ and $z$ are located along the positive real $x$-axis.)   In the present context (\ref{E:series1}) and (\ref{E:geoseries}) are replaced by
 \begin{equation}\label{E:exterior}
 F(z) = \sum\limits_{n=0}^{\infty} b_nz^{n}\ \ \ \text{and}\ \ \ 
\frac{1}{z_k - z} = \sum\limits_{n=0}^{\infty} \frac{z^n}{z_k^{n+1}}\ .
\end{equation}
Clearly, if closed-form inner products of the form $(F,\,(z_k -z)^{-1})_{{\sigma}'}$ can be found the rest of the interior SIDACKS development is straightforward.  Towards that end, replacing $f$ and $g$ by $F$ and $1/(z_k - z)$ in (\ref{E:sigg1}) and replicating the steps that led up to (\ref{E:sigg3}) yields:
\begin{equation}\label{E:Fsigg}
 (F,\,(z_k - z)^{-1}){\ls}_{{\sigma}'}\, = \,p_k\,\{F(p^{*}_k)\}^{*}\ ,
\end{equation}
where $p_k \eq 1/z_k$ as before.  Recall that a discussion of the merits of the SIDACKS procedure just outlined  and Szeg\H{o} kernel based interpolation schemes was given at the end of Section~\ref{S:SIDACKS}. 

 DIDACKS theory can be also be developed for the interior of a disk in much the same way.  Using the previous notational conventions in an obvious way, the energy inner product for the interior of a unit disk can be defined as
\begin{equation}\label{E:Einterior}
(F,\,G){\ls}_{D'} \eq \frac1{2\pi}\int\limits_{r=0\ }^{\ \,1}\int\limits_{\theta=0}^{\,2\pi} \bigg(\frac{d\,F}{d\,z}\bigg)^{*}\bigg(\frac{d\,G}{d\,z}\bigg)\, r\,d\,r\,d\,\theta\ .
\end{equation}
Now, however, $F$ and $G$ are assumed to have a power series expansion of the form
\begin{equation}\label{E:Fseries2}
F(z) = \sum\limits_{n=1}^{\infty} b_nz^{n}\ ,
\end{equation}
where the constant term ($b_0$) has been omitted from the analogous series given in (\ref{E:exterior}).  This constant term was also omitted from (\ref{E:series1}) for reasons discussed in Section~\ref{S:PreDIDACKS}.  This means that
\begin{equation}\label{E:zkzzk}
 {\zeta}_k(z) \eq \frac1{(z_k - z)} - \frac1{z_k} =
\frac{z}{z_k(z_k - z)} = \sum\limits_{n=1}^{\infty} \frac{z^n}{z_k^{n+1}}
\end{equation}
will be used as basis function here instead of $1/(z_k - z)$. 
Replacing $G$ in (\ref{E:Einterior}) with this series expansion and replicating the steps that led to (\ref{E:CollPropD}) yields the following result 
\begin{equation}\label{E:CollPropEP}
(F,\,(z - z_k)^{-1}){\ls}_{D'} = \frac{p_k^2}2\,\{F_z(p_k^{*})\}^{*}\ .
\end{equation}
Interior complex SIDACKS and DIDACKS theory can obviously be easily built upon the above results in complete analogy with the exterior treatments in Sections~\ref{S:SIDACKS} and \ref{S:DIDACKS}, thus it is not necessary to include the details here.  The \R2 development also proceeds in a completely analogous fashion to that of Section~\ref{S:RealPlane}. 

\newpage

\appendix
\begin{center}
 \begin{Large}{\textbf{Appendix A}}\end{Large}
\end{center}

\renewcommand{\theequation}{A-{\arabic{equation}}}
\setcounter{equation}{0}

\section*{\hfil Discussion of Educational Aspects\hfil}\label{S:A}
\ \hfill  \\
\vskip -.2in 

\noindent
  As noted in the introduction, there are various educational possibilities that are inherent in the general DIDACKS approach.  Since these education issues will not be explored systematically in any other venue by the author, several relevant points are summarized here.  After these points are made, this appendix ends with  a brief discussion of self-educational aspects that, as indicated in the introduction, are considered to be of equal or even greater significance than these educational ones.

  Before considering these issues it is useful to take a step back and briefly look at the bigger picture.  While much of the article has been written at an accessible level, this still does not mean that it was intended as an introductory one and, as such, it does not really stand alone in this regard (i.e., neophytes probably should have the guidance of a more seasoned intermediary before approaching much of the material here).  In order to simplify matters in general, although the author might have wished it differently, with regards to the overall level of material there is a curious attribute of DIDACKS theory itself that it has an certain innate multi-scale accessibility, which is to say that any article, even a relatively brief one (which this one is not) can, in some sense or other, cover a very large span of readership---from material that is at the introductory level to material dealing with open research problems; moreover, at both ends of the spectrum the problems and challenges posed seem to be unique and interesting.  For example, at the introductory or educational level, it is possible to contemplate combining the material of many core courses into one DIDACKS centered course, so that material that might not otherwise be covered in a core curriculum can be covered.  Related educational items are: 
 
\begin{enumerate}
\item
DIDACKS theory can serve as a means of introducing upper level undergraduate or graduate students to approximation theory, harmonic analysis, functional analysis, integral equation theory, computational linear algebra, kernel theory, Bergman kernel theory, RKHS theory and other advanced topics.
Students who have some previous exposure to complex variables will be familiar with the functional form $1/(z - z_k)$ and it is not too big a conceptual jump to consider it as a kernel within the wider context of DIDACKS theory.   In particular, RKHS theory should be considered for inclusion in introductory functional analysis courses, but it usually is not, in part, due to conceptual hurdles, and DIDACKS theory may be a way of lowering these hurdles.  One way to overcome these conceptual hurdles is to have students take a hands-on approach by having them develop thir own software implementations.  In this regard, the reproducing property of RKHS fits allows the students to validate their own software.  The basic DIDACKS approach is even more ideally suited to this style of supplemented self-instruction than RKHS theory is due to the overall accessibility of the material.
\item
Besides trying to motivate others to consider injecting RKHS theory, where appropriate, into classes, one goal here that is even more central is to motivate lecturers to present discussions of Dirichlet's principle and of the associated Dirichlet form (or integral).  Others have suggested that Dirichlet's principle should be a main part of various introductory mathematics courses due to its central role in the history of various branches of analysis.  One existing significant realization of this plan is \cite{Dacorogna}, which covers real analysis via variational principles.  The material covered here is another way to inject Dirichlet's integral and Dirichlet's principle into the coverage of introductory material.
\item
Aside from the natural way that other areas of mathematics can be integrated into a consideration of the DIDACKS approach, there are also a very wide range of tangentially related scientific and engineering disciplines that it can be applied to or adapted to in a non-trivial way.  By this means applied mathematicians, or other students in question, can be given a very broad exposure to diverse areas.
\item
There are undoubtedly pedagogical advantages to taking a single-topic or focused approach to the coverage of basic material.  One advantage is the possibly of an increased level of interest.   A second is that students often mentally retain mathematical tools by distinct separate categories, which are linked to only specific subjects, and thus miss common mathematical themes; hence, studying topics from the perspective of a different coherent mathematical framework could help prevent this.  Along these same lines, with a coherent theme it is reasonable to expect that students will  do a better job of integrating mathematical material with applications and thus retain it better.  From a research point of view, these same attributes clearly confer cross-fertilization possibilities.
\item
 In connection with point 1 above, it is also possible to use the DIDACKS approach as a framework for a self-instruction courses in scientific or applied programming courses where numerical consistency checks are extensively utilized.  For example, in the complex setting, for some specified $f$ by taking finite numerical differences along different directions and at different points, the fact that $d\,f/d\,z$ is both computed correctly and that $f$ is analytic can be easily directly spot-checked by the student programmer.  While the primary implementation tool for overall consistency checks is the replication property associated with the DIDACKS interpolation fits themselves, it should be noted that symmetric configurations of interpolation point positions in both the complex and real plane may well produce a consistent collocation check result even though implementation errors are present due to the existence of various underlying reflection and symmetry principles.  Thus a software consistency check should entail using not only irregular interpolation point locations, but also an $f$ that varies in value from one interpolation point to the next.  
\item
 Section~{\ref{S:SIDACKS}} and Section~{\ref{S:DIDACKS}} are largely self contained. Since the material in Section~{\ref{S:SIDACKS}} up through (\ref{E:fitform2}) does not require a significant functional analysis background and is accessible at a very low level, when placed in a suitable context it can be used as a stepping stone to more advanced topics; moreover, it is possible to design a course around the content of Sections~{\ref{S:SIDACKS}} and {\ref{S:DIDACKS}} that introduces a variety of pure and applied concepts.   Furthermore, after closer examination of the content of this article, it should be apparent that much of the material often present in an introductory course in complex analysis has been covered and that those topics that have not been covered can be introduced with a little ingenuity.  For example, one topic not covered is conformal mapping, but it is obvious that this topic can be naturally integrated by introducing pre-mappings from other regions to the unit disk (and then return post-mappings).  (Other classes of functions besides analytic functions can even be considered in this connection provided that the class of mappings entertained here is sufficiently broad.)  One other common topic not directly dealt with in the body of the paper is residue theory---see the next item.
\item
Those students who are already knowledgeable in the theory of residues and poles, as well as some others, may slip into a momentary lapse of confusion and think that since residue theory can be used to evaluate isolated poles at known positions by performing a simple closed line integral, this process can be used here too and thus that the formalism developed herein is, in some sense, superfluous.  To see the issue here, consider the case where the functions in question, $f(z)$, are analytic in the exterior region and all their poles are located in the interior region at unknown locations.  The point, of course, is that it is assumed that noting is known even in principle about the values of $f(z)$ for $|z| < 1$ and all that is known in this interior region is the location of the poles that are to be used in the approximating form; hence, any part of a line integral of $f$ that is inside the unit disk is unknown.  (As discussed below, $f$ may well arise from a continuous source distribution rather than from a denumerable set of isolated poles.)  By directly trying to use the theory of residues under these restrictions the student can convince him or herself that residue theory cannot otherwise supply a sufficient amount of information to uniquely determine the pole strengths.

  From another perspective, consider the question as to whether some combination of simple poles in the interior of a disk exists that can produce a null result in the exterior region: $f(z) = 0$ for $|z| \geq 1$.  This question cannot be directly resolved by using line integrals in the exterior region, but it is answered in the main body of the text by other techniques. 
\item
While it may not at first be obvious, with a sufficient application of ingenuity numerous and varied interesting problem sets can be generated to accompany lecture notes based on the DIDACKS approach.  One example here will have to suffice:  Find a polynomial expression in terms of $u(\theta) = e^{i\theta}$ for the extremals of $|\varphi|$ [given by (\ref{E:fitform})] on the unit circle.  (Notice that for $N_k < 4$ closed-form expressions can be written out for the pole strengths.)  Solution:  (a) Take $d\ /d\,\theta$ of ${\varphi}^*\varphi$ and set the result to zero.  (b) Substitute $1/u$ for $u^*$.  (c) Multiply through by $u$ and then by the common denominator.
\item
A thorough discussion of the interplay of power series and analytic functions is rightly considered to be a core part of standard complex variable courses. 
In conjunction with the precise minimal requirements here for series (\ref{E:series1}), or of the corresponding smoothness requirements for harmonic functions, interested readers are encouraged to consult standard complex variable or harmonic analysis texts \cite{AxlerEtAll} and rethink the associated DIDACKS requirements for themselves.  With regards to the discussion of these particular topics elsewhere in this article, a certain amount of looseness was present so that instructors who choose to present the associated material can discuss smoothness and power series related topics to the level of detail and rigor that they wish.  One consideration here that is important for DIDACKS theory is that continuous source distributions can be entertained, which brings in concepts not usually addressed at the introductory complex variable level since it involves issues that are not raised when only denumerable sets of simple poles are considered.

 If lecturers do bring up the topics of power series in this context, certain obvious examples should also probably be included in the discussion, such as the $\mathbb{R}^2$ harmonic problem for the exterior of a unit disk ($\Omega$) where $f = 1$ on the upper half of the unit circle and $f = 0$ along the lower half of the unit circle.  (The point here is that although the Poisson integral gives a harmonic solution $f$, which is bounded, it is not altogether obvious that $\text{D}[f,\,f]$ itself is bounded.)  To further explore this example the series solution obtained by standard boundary value techniques can be used to explicitly evaluate $\text{D}[f,\,f,\,1,\,\Omega]$. 
\end{enumerate}

\begin{center}
\ \\
 \underline{\textbf{Autodidactic Aspects of Notation and Nomenclature}}\\
\ \\
\end{center}
The ability to engage in self-education has long been recognized as a fundamental trait of successful researchers.  This is partially due to the fact that new ideas often come from exposure to other concepts from fields outside of ones narrow specialty.  It was thus with some hope of alluding to certain autodidactic possibilities and also of certain cross-fertilization possibilities that the acronym DIDACKS was chosen. 

  Unless a researcher who wants to broaden his or her horizons is lucky enough to have a colleague to interact with who is familiar with some outside area of interest, there are often significant natural barriers, in and of themselves, that must overcome in order to master this other area of expertise.  By employing an overall accessible level of presentation some of the inherent natural boundaries erected by overspecialization can be overcome.  Whether the present article is in any sense successful or not on this score, this factor is at least brought up here for general consideration by others.  Finally, along similar lines, a much more than usual amount of attention has been given here to notation in the hopes that it might foster more consideration and discussion in the open literature of this particular topic by others.

\newpage

\appendix
\begin{center}
 \begin{Large}{\textbf{Appendix B}}\end{Large}
\end{center}

\renewcommand{\theequation}{B-{\arabic{equation}}}
\setcounter{equation}{0}

\section*{\hfil Dirichlet Integrals for Analytic or Antianalytic Functions\hfil}\label{S:B}
\ \hfill  \\
\vskip -.2in 

\noindent
   
 In this appendix complex valued functions are consider to be either analytic or antianalytic. Such functions are a subset of those that depend on both $z$ and $\bar{z}$.  The assertion that $f$ is analytic means that $\partial f/\partial \bar{z} = 0$ or that $d\,f/d\,{z}= \partial f/\partial \bar{z}$.  Likewise, the assertion that $f$ is antianalytic means that $\partial f/\partial {z} = 0$.   After a little thought it might seem obvious that the correct generalization of the Dirichlet integral 
\begin{equation}\label{E:B0}
\text{D}_c[f,\,g] \eq \iint\limits_{\Omega}\ \left(\frac{d\,f}{d\,z}\right)^*\left(\frac{d\,g}{d\,z}\right)\ d\,A
\end{equation}  
 in the case where either $\partial f/\partial \bar{z} = 0$ or $\partial f/\partial {z} = 0$ is the expression
\begin{equation}\label{E:B2}
\mathbb{D}_c[f,\,g] \eq \iint\limits_{\Omega}\  \Blbrac\left(\frac{\partial\,f}{\partial\,z}\right)^*\left(\frac{\partial\,g}{\partial\,z}\right) + \left(\frac{\partial\,f}{\partial\,\bar{z}}\right)^*\left(\frac{\partial\,g}{\partial\,\bar{z}}\right)\Brbrac \ d\,A\ ,
\end{equation}  
where a different type face has been used for the new form for the Dirichlet integral.  Let 
\begin{equation}\label{E:B3}
\mathscr{J}_c[f,\,g] \eq \Blbrac\left(\frac{\partial\,f}{\partial\,z}\right)^*\left(\frac{\partial\,g}{\partial\,z}\right) + \left(\frac{\partial\,f}{\partial\,\bar{z}}\right)^*\left(\frac{\partial\,g}{\partial\,\bar{z}}\right)\Brbrac
\end{equation}  
denote the integrand here.  The question addressed in this appendix is how to justify this form of $\mathscr{J}_c[f,\,g]$ staring from expressions that are obviously related to $\mathbb{R}^2$ Dirichlet forms.

  Consider the expression
\begin{equation}\label{E:B4}
\mathscr{F}_c[f,\,g] \eq \Blbrac\left(\frac{\partial\,f}{\partial\,x}\right)^*\left(\frac{\partial\,g}{\partial\,x}\right) + \left(\frac{\partial\,f}{\partial\,y}\right)^*\left(\frac{\partial\,g}{\partial\,y}\right)\Brbrac\ .
\end{equation}
For $g = f = u + iv$, (for general $C^\infty$ $u$ and $v$) it is easy to show that
\begin{equation}\label{E:B5}
\mathscr{F}_c[f,\,f] \eq \left(\frac{\partial\,u}{\partial\,x}\right)^2 + \left(\frac{\partial\,v}{\partial\,x}\right)^2 + \left(\frac{\partial\,u}{\partial\,y}\right)^2 + \left(\frac{\partial\,v}{\partial\,y}\right)^2 
\end{equation}
or that 
\begin{equation}\label{E:B6}
\iint\limits_{\Omega}\ \mathscr{F}_c[f,\,f] \ d\,A\  = \text{D}[u,\,u] + \text{D}[v,\,v]\,,
\end{equation}
hence $\mathscr{F}_c[f,\,g]/2$ is a good candidate to examine as the correct integrand for the more general form of the complex Dirichlet integral.  Since $\partial\, z/\partial\,x = \partial\, \bar{z}/\partial\,x = 1$ and $\partial\, z/\partial\,y = -\partial\,\bar{z}/\partial\,y = i$, the chain rule gives
\begin{equation}\label{E:B7}
\frac{\partial\,f}{\partial\,x} = \frac{\partial\,f}{\partial\,z} + \frac{\partial\,f}{\partial\,\bar{z}}\ \ \ \text{and}\ \ \ \frac{\partial\,f}{\partial\,y} = i\frac{\partial\,f}{\partial\,z} - i\frac{\partial\,f}{\partial\,\bar{z}}\ .
\end{equation}
A straightforward evaluation thus gives
\begin{equation}\label{E:B9}
\mathscr{F}_c[f,\,g]/2 = \mathscr{J}_c[f,\,g]
\end{equation}
as expected, which justifies definition (\ref{E:B2}) from a different perspective.

\newpage


\end{document}